\documentclass[12pt]{article}

\usepackage{amssymb}
\usepackage{graphicx}
\usepackage{cite}

\newcommand{\const}{\,{\rm const}\,}

\def\be{\begin{equation}}
\def\ee{\end{equation}}
\def\bea{\begin{eqnarray}}
\def\eea{\end{eqnarray}}

\title{\bf{A classification of near-horizon geometries of extremal vacuum black holes}}

\author{Hari K. Kunduri$^a$\footnote{h.k.kunduri@damtp.cam.ac.uk } \  and James Lucietti$^b$\footnote{james.lucietti@durham.ac.uk } \\ \\
\small \sl $^a$DAMTP, University of Cambridge \\
\small \sl Centre for Mathematical Sciences, Wilbeforce Road, Cambridge, CB3 0WA, UK
\\ \\ \small \sl $^b$Centre for Particle Theory, Department of Mathematical Sciences, \\ \small \sl University of Durham, South Road, Durham, DH1 3LE, UK
 }

\date{}

\begin{document}

\begin{titlepage}
\maketitle

\begin{picture}(0,0)(0,0)
\put(350, 300){DCPT-08/35}
\put(350, 285){DAMTP-2008-50}
\end{picture}

\begin{abstract}

We consider the near-horizon geometries of extremal, rotating black
hole solutions of the vacuum Einstein equations, including a
negative cosmological constant, in four and five dimensions. We
assume the existence of one rotational symmetry in 4d, two
commuting rotational symmetries in 5d and in both cases non-toroidal horizon topology. In 4d we determine the most
general near-horizon geometry of such a black hole, and prove it is the same as the near-horizon limit
of the extremal Kerr-$AdS_4$ black hole. In 5d, without a
cosmological constant, we determine all possible near-horizon
geometries of such black holes. We prove that the only possibilities are one family
with a topologically $S^1 \times S^2$ horizon and two distinct
families with topologically $S^3$ horizons. The $S^1 \times S^2$
family contains the near-horizon limit of the boosted extremal Kerr
string and the extremal vacuum black ring. The first topologically
spherical case is identical to the near-horizon limit of two
different black hole solutions: the extremal Myers-Perry black hole
and the slowly rotating extremal Kaluza-Klein (KK) black hole. The
second topologically spherical case contains the near-horizon limit
of the fast rotating extremal KK black hole. Finally, in 5d with a
negative cosmological constant, we reduce the problem to solving a
sixth-order non-linear ODE of one function. This allows us to
recover the near-horizon limit of the known, topologically $S^3$,
extremal rotating $AdS_5$ black hole. Further, we construct an
approximate solution corresponding to the near-horizon geometry of a
small, extremal $AdS_5$ black ring.

\end{abstract} \thispagestyle{empty} \setcounter{page}{0}
\end{titlepage}

\tableofcontents
\newpage

\section{Introduction}
Asymptotically flat and anti de Sitter (AdS) black hole solutions in four and five dimensions are of interest in the context of string theory and AdS/CFT respectively, as they provide an effective description of the strong coupling dynamics in certain sectors of the dual conformal field theories. Focusing on supersymmetric states often allows one to evade the problem of performing computations at strong coupling, as such states tend to be protected. This provides the opportunity to reproduce the Hawking-Bekenstein entropy of the black hole in question from a microstate counting in the weakly coupled field theory.

In recent years, great progress has been made in the
construction of supersymmetric black holes both in ungauged
supergravity~\cite{ Gauntlett:2002nw, ReallNH, Gaiotto:2005gf,
Elvang:2004rt, Gauntlett:2004wh, Elvang:2004ds, Gauntlett:2004qy,
Elvang:2005sa, Bena:2005ni} and gauged
supergravity~\cite{Gutowski:2004ez, Gutowski:2004yv, Chong,
Kunduri:2006ek, her, Kunduri:2006uh, Kunduri:2007qy}, largely due to
systematic classification techniques available for BPS
solutions~\cite{Gauntlett:2002nw, Gauntlett:2003fk,
Gutowski:2004yv}. As is well known, supersymmetric black holes are necessarily
extremal. Curiously, recent work on the attractor mechanism
(see \cite{Sen:2007qy} for a comprehensive review) has revealed that
in fact it may be extremality rather than supersymmetry which is
responsible for the success of the entropy counting of black holes
in flat space~\cite{Astefanesei, Dabholkar}. In the case of extremal
but non-supersymmetric black holes, the attractor mechanism was
established upon the assumption that the near-horizon geometry of an
extremal black hole must have an $SO(2,1)$ symmetry \cite{Sen2}.
This assertion was proved in four and five dimensions in~\cite{KLR},
in a large class of theories, under the assumption that the black
hole is axisymmetric in four dimensions and has two commuting
rotational Killing vector fields in five dimensions (see
also {\cite{FKLR} for generalisations for $D>5$).\footnote{Note
that it has been proved that a stationary, non-extremal black hole in all dimensions is necessarily
axisymmetric~\cite{HIW, IM}, i.e. has at least one rotational Killing vector field so the total symmetry is at least $R \times U(1)$.} Indeed, there has
been recent success in counting the microstates of extremal,
non-supersymmetric black holes in four and five
dimensions~\cite{EH1, EM, Horowitz:2007, Reall:2007jv,
Emparan:2008qn}.

The classification problem of stationary black holes in higher
dimensions is also of intrinsic interest\footnote{Note that even in
four dimensions, there is no uniqueness theorem for asymptotically
$AdS_{4}$ black holes.}. From this point of view supersymmetry is
merely a technical tool allowing one to study the classification
problem in a more constrained setting. Similarly, extremality may
also be used as a simplifying assumption. This is because any
extremal black hole admits a near-horizon limit, a geometry in its
own right which solves the same field equations~\cite{ReallNH, KLR}.
The advantage of this is that determining and thus classifying
near-horizon geometries is a technically simpler problem: it becomes
a $D-2$ dimensional problem of Riemannian geometry on a compact
space (i.e. spatial sections of the horizon). Given a classification
of near-horizon geometries in some theory, one can deduce certain
information about what black hole solutions are allowed. In
particular it can allow one to rule out the existence of extremal
black holes with a certain horizon topology. Furthermore, this
analysis determines not only the possible horizon topologies, but
also determines their geometry explicitly.  The one
disadvantage of this method is that the existence of a near-horizon
geometry does not guarantee the existence of an extremal black hole
solution with that near-horizon geometry.

Previously, certain classifications of near-horizon geometries have been achieved in a variety of ungauged supergravities~\cite{ReallNH,Gutowski:2004bj, GMR, CRT0}, where the combined use of supersymmetry and the near-horizon limit is particularly fruitful. The main success of this is it allowed the proof of a uniqueness theorem for asymptotically flat, topologically spherical, superysymmetric black holes in five dimensional ungauged supergravity: the only solution turns out to be BMPV~\cite{ReallNH, Gutowski:2004bj}. In the gauged case the near-horizon equations are more complicated and a classification of near-horizon geometries was achieved using an extra assumption: the black hole admits two commuting rotational symmetries~\cite{Kunduri:2006uh, Kunduri:2007qy}. This ruled out the existence of supersymmetric AdS$_5$ black rings with these symmetries.

In this work, we consider the classification of near-horizon
geometries in a setting without supersymmetry in four and five
dimensions. For simplicity we will consider near-horizon geometries
of extremal black hole solutions to Einstein's vacuum equations and
allow for a negative cosmological constant. As a result, we can consider asymptotically flat (and KK in 5d) and AdS black
holes respectively.\footnote{Our analysis also covers multi-black
holes in the sense that one can take a near-horizon limit with
respect to each connected component of the horizon -- in this limit
the effects of the other components of the horizon is lost
(see~\cite{Gauntlett:2004wh} for a supersymmetric example).} In the
pure vacuum in five dimensions there are a number of known examples of
extremal black holes and their associated near-horizon geometries:
the extremal boosted Kerr string, the extremal Myers-Perry black
hole~\cite{MP}, the extremal black ring~\cite{PS}, and two different
extremal limits of the KK black hole~\cite{Rasheed, Larsen} (often
termed ``slow'' and ``fast'' rotating)\footnote{These correspond to
$G_4J<PQ$ and $G_4J>PQ$. The solution with $G_4J=PQ$ is nakedly
singular.}. In contrast, in the presence of a negative cosmological
constant only one example is known: the extremal limit of the topologically spherical, rotating $AdS_{5}$ black hole found in~\cite{HHT}. Indeed, an interesting open question
concerns the existence of asymptotically AdS$_5$ black rings. No
such solutions are currently known. Furthermore, the systematic
solution generating techniques available for vacuum
gravity~\cite{BZ1,BZ2}, are not available in the presence of a
cosmological constant. Thus it appears that a near-horizon analysis
is one of the few systematic techniques available to obtain
information on the existence of AdS$_5$ black rings (at least in the
extremal sector).

We use the assumption of axisymmetry in 4d and two commuting
rotational symmetries in 5d, which means the near-horizon geometry
is cohomogeneity-1 in both cases; therefore everything reduces to
ODEs. Our analysis will employ both local and global considerations
(i.e. compactness of spatial sections of the horizon). The global
arguments allow one to avoid solving the differential equations
generally, thus simplifying the problem. The main results of this
paper may now be stated.

\paragraph{Theorem 1} Consider a four-dimensional non-static and axisymetric near-horizon geometry, with a compact horizon section of non-toroidal topology, satisfying $R_{\mu\nu}=\Lambda g_{\mu\nu}$ for $\Lambda \leq 0$. If $\Lambda=0$ then is must be the near-horizon limit of the extremal Kerr black hole. If $\Lambda<0$ it must be the near-horizon limit of the extremal Kerr-$AdS_{4}$ black hole.

\paragraph{Remarks:}
\begin{itemize}
\item For $\Lambda=0$ the same result has been proven in~\cite{Haj}, and again in the context of isolated
horizons in~\cite{LP}. Their analysis included a Maxwell field (in
which case the result is the near-horizon geometry of extremal
Kerr-Newman).\footnote{Since the first version of our paper, some results have also been derived without the assumption of axisymmetry~\cite{J}.}
\item Static near-horizon geometries of this form have been considered in~\cite{CRT}. For $\Lambda=0$ it was shown that the near-horizon geometry is a direct product of 2d Minkowski space and a flat $T^2$. However, in the context of black holes this may be excluded by the horizon topology theorems~\cite{CW,GSWW}. For $\Lambda<0$ it was shown that it is a direct product of $AdS_2$ and a compact Einstein space of negative curvature: this is incompatible with our assumption of axisymmetry.
\item  Topological censorship~\cite{GSWW} implies that for asymptotically flat and globally AdS$_4$ black holes the horizon section can not have toroidal topology. Thus our result implies that the near-horizon geometry of any asymptotically flat (or globally AdS$_4$), Ricci flat (or negative curvature Einstein-space), stationary and axisymmetric extremal black hole is given by the near-horizon limit of Kerr (or Kerr-AdS$_4$).
\item Note that for four dimensional {\it non-extremal} rotating black holes, axisymmetry has been proved to be a consequence of stationarity (even in AdS~\cite{HIW}). Therefore it is reasonable to expect the same to occur for extremal black holes and thus their near-horizon limits.\footnote{In fact after our paper first appeared it has been shown that a 4d stationary, extremal rotating black hole must be axisymmetric~\cite{HI}.}

\end{itemize}

\paragraph{Theorem 2} Consider a five-dimensional non-static near-horizon
geometry, with a compact horizon section $\mathcal{H}$ of non-toroidal topology, and a
$U(1)^2$ isometry group with space-like orbits, satisfying
$R_{\mu\nu}=0$. Then it must be contained in one of the following three families: a three parameter family with  $\mathcal{H}=S^1 \times S^2$, given by equation
(\ref{ring}); a two parameter family with $\mathcal{H}=S^3$ (case A) given by
(\ref{S3A1})-(\ref{S3A2}); a three parameter family with $\mathcal{H}=S^3$ (case B) given by
(\ref{S3B}). See the main results section (\ref{Summary5d}) for more
details and explicit metrics.

\paragraph{Remarks:}
\begin{itemize}
\item Static vacuum near-horizon geometries were considered in~\cite{CRT}. It was shown that they must be the direct product of 2d Minkowski space and a flat compact 3d space. However, in the context of black holes, these may be ruled out by the black hole horizon topology theorem~\cite{GS, Galloway}.
\item $S^1\times S^2$ case: In a region of parameter space it is isometric to the near-horizon limit of extremal boosted Kerr string. Further, for a particular value of the boost parameter (i.e. such that the string is tensionless) it is isometric to the near-horizon limit of the asymptotically flat extremal vacuum black ring~\cite{KLR}.
\item $S^3$ case A: This is isometric to the near-horizon limit of two different black holes: extremal Myers-Perry (which must have two non-zero angular momenta $J_i$) and the slow rotating extremal KK black hole ($G_4J<PQ$). In a special case (corresponding to $J_1=\pm J_2$ and $J=0$ respectively) the rotational symmetry group enhances to $SU(2)\times U(1)$ (or $SO(3)\times U(1)$) and the near-horizon geometry is a homogeneous space.
\item $S^3$ case B: In a region of parameter space it is isometric to the near-horizon limit of the fast rotating extremal KK black hole ($G_4J>PQ$). This solution always has total rotational symmetry group $U(1)^2$ (i.e it never gets enhanced as in case 2) .
\item Any extremal vacuum black hole in five dimensions with $R\times U(1)^2$ isometry group and compact horizon sections of non-toroidal topology, must have a near-horizon geometry contained in one of the three families
above. Note that toroidal horizon topology is not allowed by the
black hole topology theorems of~\cite{GS, Galloway}.
\item For a non-extremal rotating black hole in five dimensions, it has been proved that stationarity implies the existence of one rotational symmetry~\cite{HIW, IM}. Therefore one expects extremal black holes to also have one rotational symmetry. We have assumed two rotational symmetries, a property satisfied by all known black hole solutions in five dimensions, although there is no general argument for this.
\end{itemize}

We have not been able to determine all possible near-horizon geometries with two rotational symmetries and compact horizons in 5d with a negative cosmological constant. We have reduced the problem to solving one 6th order ODE of one function. The only family of solutions to this ODE we know of corresponds to the 2-parameter family of near-horizon geometries of the extremal rotating AdS$_5$ black holes of~\cite{HHT}. If a vacuum extremal AdS black ring with two rotational symmetries does indeed exist it must correspond to a solution to our ODE. An AdS black ring would possess a number of length scales: $R_1$ the radius of the $S^1$ of the horizon, $R_2$ the radius of the $S^2$ of the horizon and $\ell$ the AdS length scale. A small AdS black ring would be one such that $R_1<<\ell$ and $R_2 <<\ell$. In this regime the black ring would not ``see'' the effects of the AdS boundary conditions and one would expect it to be well approximated by an asymptotically flat black ring. Therefore, by perturbing about the solution corresponding to the near-horizon of the asymptotically flat black ring, one should be able to construct a first order correction (valid for small $R_i/ \ell$) representing the near-horizon of a small extremal AdS ring. We have performed this calculation and find that there exist regular perturbations which preserve the $S^1 \times S^2$ topology of the horizon. It is tempting to conclude that this provides some evidence for the existence of, at least a small, extremal vacuum black ring in AdS$_5$.

The organisation of this paper is as follows. In Section 2 we present a self-contained summary of our main results. Section 3 provides a review of general features of near-horizon geometries with rotational symmetries and we present the field equations to be analysed. Section 4 deals with the four dimensional case, including a negative cosmological constant. In Section 5 we consider five dimensional near-horizon geometries: first we examine the general case (including a negative cosmological constant), then turn to a classification of all solutions in the pure vacuum case and finally we investigate the existence of solutions describing the near-horizon limit of an extremal black ring in $AdS_{5}$. Section 6 concludes with a discussion of our results. The details of various technical results used throughout the paper are given in the Appendices.

\section{Summary of main results}
In this section we will state more explicitly the main results of this paper. This section is intended to be a self-contained summary without derivations; these are provided in the rest of the paper.

\subsection{Vacuum near-horizon geometries in $D=4$ including a negative cosmological constant}
Consider a 4d stationary, axisymmetric extremal black hole, with a compact horizon section of non-toroidal topology, satisfying $R_{\mu\nu}=\Lambda g_{\mu\nu}$ with $\Lambda \leq 0$. We have proved that its near-horizon limit must be given by

\be ds^2= \Gamma(\sigma) [ -C^2r^2 dv^2+2dvdr]
+\frac{\Gamma(\sigma)}{Q(\sigma)}d\sigma^2 +
\frac{Q(\sigma)}{\Gamma(\sigma)} (dx+rdv)^2 \ee where \be
\Gamma=\beta^{-1}+ \frac{\beta \sigma^2}{4}, \qquad Q = -\frac{\beta
\Lambda}{12}\sigma^4 - (C^2 + 2\Lambda\beta^{-1})\sigma^2 +
4\beta^{-3}(C^2\beta + \Lambda) \ee and $C>0$, $\beta>0$ are
constants. $Q$ must have four distinct real roots. The coordinate
ranges are given by $\sigma_1\leq \sigma \leq \sigma_2$ where
$\sigma_2$ is the smallest positive root of $Q$ and
$\sigma_2=-\sigma_1$ and $x \sim x+ 2\pi k$ where
$k=\Gamma(\sigma_2)/(C^2\sigma_2)$. This is actually a 1-parameter
family of solutions due to a scaling symmetry of the solution which
allows one to set $C^2$ or $\beta$ to any desired value. It has
isometry group $SO(2,1) \times U(1)$ with the orbits given by circle
bundles over $AdS_2$ and its cohomogeneity-1. The horizon is at $r=0$ and spatial sections of this are $S^2$ endowed with a
cohomogeneity-1 metric. This near-horizon geometry is isometric to
that of extremal Kerr ($\Lambda=0$) or extremal Kerr-AdS$_4$
($\Lambda<0$).

A consequence of the above result is that any stationary axisymmetric extremal black hole solution (with $S^2$ horizon sections) satisfying $R_{\mu\nu}=\Lambda g_{\mu\nu}$ for $\Lambda \leq 0$ must have a near-horizon geometry given by that of extremal Kerr ($\Lambda=0$) and  Kerr-AdS$_4$ ($\Lambda<0$).

\subsection{Vacuum near-horizon geometries in $D=5$}
\label{Summary5d}
Consider a 5d Ricci flat extremal black hole with $R\times U(1)^2$ symmetry (i.e. stationary plus two rotational symmetries) and assume spatial sections of the horizon are not toroidal. We have proven that its near-horizon geometry must be contained in one of three families:

\paragraph{$S^1 \times S^2$ horizon} The near-horizon geometry in this case can be written as
\bea
ds^2 &=&C^2a^2(1+\sigma^2)\left[- C^2r^2 dv^2+2dvdr \right] \nonumber \\ &&+ \frac{a^2(1+\sigma^2)}{1-\sigma^2} d\sigma^2 +
\frac{4a^2(1-\sigma^2)}{(1+\sigma^2)} \left(d\phi + \Omega dx^2 +C^2rdv \right)^2+ \frac{1}{4C^4a^2}(dx^2)^2 \label{ring}
\eea
where $-1 \leq \sigma \leq 1$, $\phi \sim \phi+2\pi$ and $x^2 \sim x^2+L$. The solution is parameterized by the constants $(C,a,\Omega ,L)$ where $C,a,L>0$, however due to a scaling symmetry one of $C,\Omega,L$ may be set to any convenient value. It is therefore a three parameter family of solutions. The isometry group of this geometry is $SO(2,1) \times U(1)^2$. The orbits of $SO(2,1)$ are circle bundles over $AdS_2$ and the geometry is cohomogeneity-1. The horizon is at $r=0$ and spatial sections of this are $S^1 \times S^2$ endowed with a
cohomogeneity-1 metric. In fact, the $C^2|\Omega|< 1/(4a^3)$ case is identical to the near-horizon limit of the boosted extremal Kerr-string with boost parameter $\beta$ and Kerr parameter $a$, see~\cite{KLR}. This can be seen by defining $\tanh \beta = 4a^3 C^2\Omega$ and setting $C^2=1/(2a^2 \cosh\beta)$ (which we are free to do due to the scaling symmetry mentioned).  Further if one chooses the boost such that $\sinh^2\beta=1$ it is isometric to the near-horizon limit of the extremal vacuum black ring, see~\cite{KLR}.

\paragraph{$S^3$ horizon: case A}
The main assumption of our analysis is the existence of a $U(1)^2$ rotational symmetry. As is typical of rotating solutions in 5d, in this class there is a special case in which the rotational symmetry group enhances to $SU(2)\times U(1)$. It is convenient to write this special case in a separate coordinate system.

The more symmetric case can be written as
\be
ds^2 = \Gamma[-C^2r^2 dv^2+2dvdr]+\frac{2\Gamma}{C^2}
(d\psi+\cos\theta d\phi +C^2rdv)^2+ \frac{\Gamma}{C^2}(d\theta^2+\sin^2\theta d\phi^2) \label{S3A1}
\ee
where $0 \leq \psi \leq 4\pi$, $0 \leq \phi \leq 2\pi$, $0 \leq \theta \leq \pi$ are the usual Euler angles on $S^3$. The solution is parameterized by the constants $(C^2,\Gamma)$, however due to a scaling symmetry it is a one parameter family. This solution has an isometry group $SO(2,1)\times SU(2) \times U(1)$. The orbits of $SO(2,1)$ are circle bundles over $AdS_2$ and the geometry is a homogeneous space. The horizon is at $r=0$ and spatial sections of this are $S^3$ endowed with a homogeneous metric. It turns out that this case is isometric to both the near-horizon limit of the $J_1=J_2$ extremal Myers-Perry black hole and the near-horizon limit of the $J=0$ extremal KK black hole.

The generic case is more complicated. It can be written as
\bea
ds^2 &=& \sigma [ -C^2 r^2 dv^2 +2dvdr ] \nonumber \\ &&+\frac{\sigma
d\sigma^2}{Q(\sigma)} + \left( C^2\sigma- \frac{c_2}{\sigma}
\right) \left( dx^1 + rdv+ \frac{\sqrt{-c_2c_1} \; dx^2}{C (C^2 \sigma^2 - c_2)}  \right)^2 +
\frac{Q(\sigma) (dx^2)^2}{\left( C^2 \sigma^2
-c_2 \right)} \label{S3A2}\eea
where $Q(\sigma)=-C^2 \sigma^2+c_1\sigma+c_2$ and $\sigma_1 \leq \sigma \leq \sigma_2$ where $\sigma_1,\sigma_2$ are the roots of $Q$ and $0<\sigma_1<\sigma_2$. The parameters must satisfy $c_1>0$, $c_2<0$ and $c_1^2+4C^2 c_2>0$. There is a scaling symmetry so it is really just a two parameter family of metrics. The coordinates $\phi_i$ adapted to the $U(1)^2$ rotational symmetry are defined by $\frac{\partial}{\partial \phi_i }=-d_i\left(\frac{1}{C\sigma_i}\sqrt{\frac{-c_2}{c_1}} \frac{\partial}{\partial x^1 }- \frac{\partial}{\partial x^2} \right)$ where $d_i$ are chosen so that the periods of $\phi_i$ are $2\pi$. This near-horizon geometry has an isometry group $SO(2,1) \times U(1)^2$ whose generic orbits are $T^2$ bundles over $AdS_2$ and therefore it is cohomogeneity-1 (the orbits of $SO(2,1)$ in general are line bundles over $AdS_2$). Spatial sections of the horizon $r=0$ are given by $S^3$ endowed with a cohomogeneity-1 metric. It turns out this case is isometric to the near-horizon limit of two different black holes: the extremal $J_1 \neq J_2$ Myers-Perry and the extremal $0<G_4J<PQ$ KK black hole.

In summary, this class of $S^3$ topology horizons are isometric to the near-horizon limit of either: (i) extremal Myers-Perry, (ii) slowly rotating KK black hole.

\paragraph{$S^3$ horizon: case B} In this case the near-horizon geometry is of the form
\bea
ds^2 &=& (a_2\sigma^2+a_0)[-C^2r^2 dv^2 +2dvdr] \nonumber \\ &&+  \frac{(a_2\sigma^2+a_0) d\sigma^2}{Q(\sigma)}+
\frac{2P(\sigma)}{a_2\sigma^2+a_0}\left[ dx^1 +rdv - \frac{\kappa a_2\sigma}{\alpha P(\sigma)}
dx^2 \right]^2 + \frac{Q(\sigma)}{2P(\sigma)} (dx^2)^2 \label{S3B}
\eea
where
\be
Q=-C^2\sigma^2 +c_1\sigma +c_2, \qquad P=\alpha\sigma^2+\beta \sigma+\gamma
\ee
with
\be
\alpha=-a_2(C^2a_0+a_2c_2), \qquad \beta=2c_1a_0a_2, \qquad \gamma=a_0(C^2a_0+a_2c_2)
\ee
and
\be
\kappa \equiv \sqrt{ \frac{(a_0C^2 - a_2c_2)[
c_1^2a_0a_2 +(C^2a_0+a_2c_2)^2]}{2}} \; .
\ee
The constants $(a_0,a_2,c_1,c_2,C^2)$ must satisfy $C^2a_0-a_2c_2>0$, $c_1^2a_0a_2 +(C^2a_0+a_2c_2)^2>0$ and $c_1^2+4C^2c_2>0$.\footnote{As written, the above metric is valid for $\alpha \neq 0 $. In fact the $\alpha = 0$ case is also allowed (provided $\beta\neq 0$) and may be obtained by shifting $x^1$ appropriately.} The latter condition ensures that $Q$ has two distinct real roots $\sigma_1<\sigma_2$ and the coordinate $\sigma$ must belong to the interval $\sigma_1\leq \sigma \leq \sigma_2$. This metric possesses two independent scaling symmetries and thus is really just a three parameter family. It has an isometry group $SO(2,1) \times U(1)^2$ whose generic orbits are $T^2$ bundles over $AdS_2$ and therefore it is cohomogeneity-1 (the orbits of $SO(2,1)$ are generically line bundles over $AdS_2$). The horizon is at $r=0$ and spatial sections of this are $S^3$ endowed with a cohomogeneity-1 metric. Using one of the scaling symmetries one can always set $c_1^2+4C^2c_2=4C^4$: then, the region of parameter space defined by $a_2>0$ and $4a_2C^{-2}+2C^{-4}(C^2a_0-a_2c_2)< [c_1^2a_0a_2+(C^2a_0+a_2c_2)^2]/(C^6a_2)$  can be shown to be identical to the near-horizon geometry of the fast rotating extremal KK black hole (i.e. $G_4J>PQ$).

\subsection{Vacuum near-horizon geometries in $D=5$ with a negative cosmological constant}
Consider a 5d near-horizon geometry with two commuting space-like Killing vectors which satisfies $R_{\mu\nu}=\Lambda g_{\mu\nu}$. We have shown the problem is equivalent to solving the two coupled ODEs:
\be
\frac{d^2Q}{d\sigma^2}+2C^2+6\Lambda\Gamma=0, \qquad \frac{d}{d\sigma}\left( \frac{Q^3}{\Gamma} \frac{d^3 \Gamma}{d\sigma^3} \right) -10\Lambda Q^2 \frac{d^2\Gamma}{d\sigma^2}=0
\ee
for the pair of functions $(\Gamma(\sigma),Q(\sigma))$ where $C>0$ is a constant and $\Gamma>0$. Observe that eliminating $\Gamma$ gives a 6th order non-linear ODE. The near-horizon geometry is given in coordinates $(v,r,\sigma,x^1,x^2)$ by:
\be
ds^2=\Gamma[-C^2r^2dv^2+2dvdr]+ \frac{\Gamma d\sigma^2}{Q}+ \gamma_{11}(dx^1+\omega(\sigma) dx^2)^2 + \frac{Q}{\Gamma \gamma_{11}} (dx^2)^2
\ee
with
\be
\gamma_{11}=\Gamma \frac{d}{d\sigma} \left( \frac{Q \dot{\Gamma}}{\Gamma} \right)+2C^2 \Gamma+2\Lambda \Gamma^2
\ee
and $\omega \equiv \gamma_{12}/\gamma_{11}$ is determined up to quadratures by either (\ref{omegadot1}) or (\ref{omegadot2}). Note that $\partial/\partial v$, $\partial /\partial x^1$ and $\partial /\partial x^2$ are all Killing so the metric depends on the single coordinate $\sigma$. The horizon is at $r=0$.

The most general polynomial solution to the pair of ODEs is:
\be
\Gamma=a_0+a_1\sigma, \qquad  Q=-\Lambda a_1 \sigma^3 -
(C^2+3\Lambda a_0)\sigma^2 +c_1 \sigma +c_2  \; .\ee
The resulting near-horizon geometry is a straightforward generalisation of the Ricci flat near-horizon geometry with $S^3$ horizon (case 1) in the previous section. It turns out this case (once compactness of the horizon is imposed) is exactly the near-horizon geometry of the most general known extremal rotating AdS$_5$ black hole~\cite{HHT}, which has horizon topology $S^3$.

We have not been able to find all solutions to the pair of ODEs, which prevents us from providing a classification of near-horizon geometries in this case. It would be interesting to find a solution with horizon topology $S^1 \times S^2$ thus providing a candidate for the near-horizon geometry of an extremal AdS black ring. By linearising the ODEs about the solution corresponding to the asymptotically flat black ring we have constructed an approximate solution corresponding to the near-horizon limit of a ``small'' AdS black ring, see (\ref{smallring}).

\section{Vacuum near-horizon equations}
Consider a stationary extremal black hole. In a neighbourhood of the
horizon we can introduce Gaussian null coordinates $(v,r,x^a)$,
where $V=\partial/\partial v$ is a Killing field, the horizon is at
$r=0$ and $x^a$ are coordinates on a $D-2$-dimensional spatial
section of the horizon. We will refer to this $D-2$-dimensional
manifold as $\mathcal{H}$, which we assume is orientable, compact and without a
boundary. One can take the near-horizon limit of the metric by
sending $v \to v /\epsilon$, $r \to \epsilon r$ and $\epsilon \to
0$, see~\cite{ReallNH, KLR}. This gives \be \label{GN} ds^2 =
r^2F(x)dv^2 +2dvdr+2rh_a(x)dvdx^a+ \gamma_{ab}(x)dx^a dx^b \ee where
$F,h_a,\gamma_{ab}$ are a function, one-form and Riemannian metric
on $\mathcal{H}$ which we will refer to as the near-horizon data.

In this paper we will consider the problem of determining all vacuum
near-horizon geometries allowing for a negative cosmological
constant, i.e. metrics of the form (\ref{GN}) satisfying
$R_{\mu\nu}=\Lambda g_{\mu\nu}$ with $\Lambda \leq 0$. It can be
shown that these space-time Einstein equations for the metric
(\ref{GN}) are equivalent to the following set of equations on
$\mathcal{H}$:\footnote{These two equations have appeared before,
for example in~\cite{CRT}. We have proved that all components of the
spacetime Einstein equations are satisfied if and only if
(\ref{Ric}) and (\ref{F}) are.} \bea \label{Ric}
R_{ab} &=& \frac{1}{2}h_ah_b - \nabla_{(a} h_{b)} +\Lambda \gamma_{ab}\\
F &=& \frac{1}{2}h_a h^a -\frac{1}{2}\nabla_a h^a +\Lambda \label{F}
\eea where $R_{ab}$ and $\nabla_a$ are the Ricci tensor and metric
connection of the horizon metric $\gamma_{ab}$. For later
convenience it is worth noting that using (\ref{Ric}) and (\ref{F}),
the contracted Bianchi identity for $R_{ab}$ is equivalent to the
following equation \be \nabla_a F =
Fh_a+2h^b\nabla_{[a}h_{b]}-\nabla^b \nabla_{[a}h_{b]} \label{gradF}
\; . \ee In this paper we will be concerned with solving the equations (\ref{Ric}) and (\ref{F}). Although we have not been able to solve them in general,
we will show how one can determine all solutions with a compact
$\mathcal{H}$ under extra assumptions regarding rotational
symmetries and horizon topology.

First, however, we will note a number of general implications of the
above equations for $\Lambda=0$. Observe that  \be
\int_{\mathcal{H}} R = \int_{\mathcal{H}} F = \int_{\mathcal{H}}
\frac{h^ah_a}{2} \geq 0 \ee with equality if and only if $h_a \equiv
0$. In the case $h_a \equiv 0$ it follows that $F \equiv 0$ and
$\gamma_{ab}$ is Ricci flat: the near-horizon geometry is then
simply a direct product of $R^{1,1}$ and a Ricci flat metric on
$\mathcal{H}$. In 4d and 5d this implies a flat metric on
$\mathcal{H}$. In 4d this implies $\mathcal{H}= T^{2}$, whereas in 5d there are more possibilities (which include $\mathcal{H}=T^3$)~\cite{Thurston}. Also note that in 4d
we see that the Euler number $\chi(\mathcal{H}) \geq 0$ and thus the
only possible horizon topologies are $S^2$ and $T^2$ (the latter of
which occurs only when the near-horizon geometry is a direct product
of $R^{1,1}$ and flat $T^2$). Observe that for $r>0$ the Killing
vector $V=\partial /\partial v$ cannot be timelike everywhere, i.e.
$F(x)<0$ for all $x$ is not allowed.

\subsection{Cohomogeneity-1 near-horizon geometries}
We will now restrict consideration to near-horizon geometries of stationary
extremal black holes which are axisymmetric in $D=4$ and which admit
two commuting rotational Killing vector fields in $D=5$. That is
black holes with an isometry group $R \times U(1)^{D-3}$ in $D=4,5$ whose generic orbits are $D-2$ dimensional.
Denote the generators of the $U(1)^{D-3}$ isometry by $m_i$ and
introduce coordinates adapted to these so that $m_i=
\partial /
\partial \phi^i$ with $\phi^i \sim \phi^i +2\pi $.
The near-horizon geometry inherits the $U(1)^{D-3}$ isometry group,
which implies the full near-horizon geometry is cohomogeneity-1
(see~\cite{KLR}). Therefore the near-horizon data
$(F,h_a,\gamma_{ab})$  defined on $\mathcal{H}$ are all invariant
under $m_i$.

The existence of the $U(1)^{D-3}$ isometry group restricts the
horizon topology as follows. First note that the $U(1)^{D-3}$
isometry defines an effective group action on $\mathcal{H}$,
although we do not assume it acts freely\footnote{Recall an
effective group action is defined by the following property: the
only group element which leaves all points in the manifold invariant
is the identity. A free action is one for which all the isotropy
groups are trivial.}. Thus spatial sections of the horizon,
$\mathcal{H}$, are $D-2$ dimensional closed (compact with no
boundary) and orientable manifolds with a $U(1)^{D-3}$ effective
action. For $D=4$ the only possible closed, oriented two-manifolds
admitting an effective $U(1)$ action are $S^2$ and $T^2$. In $D=5$
the only possible closed oriented 3-manifolds which admit an
effective $U(1)^2$ action are: $T^3$, $S^1 \times S^2$ and $S^3$ (as
well as the Lens spaces $L(p,q)$ which occur as it's quotients by
discrete isometry subgroups), see e.g.~\cite{Chrusciel2rot} and
references therein. Note that only in the $T^3$ case is the action
free -- in the other cases there are fixed points. In fact there are
global parts of our analysis which do not apply to the $T^3$ case
and therefore we assume non-toroidal topology in $D=4,5$ henceforth.
This does not represent much of a restriction though, as in view of
the black hole topology theorems~\cite{GSWW, GS, Galloway}, one is
mainly interested in $S^2$ topology in $D=4$ and $S^3, S^1\times
S^2$ in $D=5$.  Let us now introduce some globally defined
quantities which are central to our analysis.

Define the 1-form $\Sigma= -i_{m_1}\cdots i_{m_{D-3}}
\epsilon_{D-2}$ where $\epsilon_{D-2}$ is the volume form associated
to the metric $\gamma_{ab}$ on $\mathcal{H}$. Since the $m_i$ are
Killing fields it follows that $d\Sigma=0$. Therefore if
$H^1(\mathcal{H})=0$ there exists a globally defined function
$\sigma$ such that $\Sigma=d\sigma$. For $\mathcal{H}=S^2, S^3$ this
is the case\footnote{In fact $\sigma$ is well defined even for Lens
space topology. This is because on $S^3$ the function $\sigma$ is
invariant under the $U(1)^2$ isometries and thus it is well defined
on the quotient spaces $L(p,q)=S^3/Z_p$ where $Z_p$ is a subgroup of
the isometries. In fact all our results for $S^3$ also apply to Lens
spaces.}. For $\mathcal{H}=S^1 \times S^2$ we can argue that the
1-form $\Sigma$ is a 1-form on the quotient space
$\mathcal{H}/S^1=S^2$ as follows. Since $\Sigma \cdot m_i=0$, we
simply need to show that some combination of the Killing fields
$m_i$ are a vector field on the $S^1$. But this must be the case, as
otherwise restricting to $S^2$ one would have a metric on $S^2$ with
$U(1)^2$ isometry which is impossible. Hence as claimed, we have
shown that even in the $S^1 \times S^2$ case the closed 1-form
$\Sigma$ must be globally exact (as it is also a closed 1-form on
the $S^2$). To summarise, we have shown that for the cases of
interest $\mathcal{H}=S^2, S^3, S^1 \times S^2$ one may define a
globally defined function $\sigma$ as above; notice that $\sigma$ is
invariant under the $U(1)^2$ isometries since $m_i \cdot d\sigma=0$
and also that $d\sigma=0$ at the fixed points of the rotational
Killing fields.

In fact the fixed points of the rotational Killing fields may be
used to distinguish these topologies~\cite{Gowdy}. For $S^2, S^1
\times S^2$ there is a unique Killing field which has fixed points
and further it only has two fixed points (the poles of the $S^2$).
For $S^3$ again there are only two points where one (or a
combination of) the Killing fields vanish, however it is a different
Killing field which vanishes at each of these points. From the
definition of $\sigma$ we deduce that $d\sigma=0$ only at fixed
points of the $m_i$; it follows that for the topologies under
consideration $d\sigma$ vanishes only at these two points.

The 1-form $h_a$ on $\mathcal{H}$, which is part of the near-horizon data, can be decomposed {\it globally} using the Hodge decomposition theorem as follows:
\be
\label{globalhrep}
h= \beta +d\lambda
\ee
where $\beta$ is a co-closed 1-form\footnote{Notice that if $H^1(\mathcal{H})=0$ then $\beta$ is also co-exact.} (i.e. $\nabla_a \beta^a=0$) and $\lambda$ is a function on $\mathcal{H}$. Since $L_{m_i}h=0$ it follows that $L_{m_i}\beta=0$ and $L_{m_i}d\lambda=0$ separately. This follows from uniqueness of the Hodge decomposition together with the facts that $L_{m_i}\beta$ is co-closed (using the properties that $\beta$ is co-closed and $m_i$ are Killing fields) and also that $L_{m_i} d\lambda=d( m_i \cdot d\lambda)$ is exact. This also shows that $m_i \cdot d\lambda =c_i$ where $c_i$ are constants. However, since $\lambda$ is a function on $\mathcal{H}$ it must be periodic in $\phi^i$ where $m_i =\partial /\partial \phi^i$ -- this implies that $c_i=0$ and hence $m_i \cdot d\lambda =0$. We find it convenient to introduce the invariant function $\Gamma=e^{-\lambda}$ which satisfies $\Gamma>0$ on all $\mathcal{H}$.

We now introduce local coordinates $(\rho, x^i)$ on a horizon section which are adapted to the $U(1)^{D-3}$ isometry: \be \label{hor}
\gamma_{ab} dx^a dx^b = d\rho^2+\gamma_{ij}(\rho) dx^i dx^j \ee
where $\partial /
\partial x^i $ are Killing fields and $i=1,\cdots, D-3$.  Observe
that in $D=4$ it is necessarily the case that $m_1 \propto \partial
/
\partial x^1$, whereas in $D=5$ the $\partial /\partial x^i$ can be
linear combinations of the $m_i$ and thus need not have closed
orbits. In these coordinates all scalar invariants only depend on
$\rho$, so for example $\Gamma=\Gamma(\rho)$ and $\sigma=\sigma(\rho)$. Now consider the co-closed 1-form $\beta$ above. The fact that it is invariant under the $m_i$ means $\beta=\beta_{\rho}(\rho)d\rho+\beta_i(\rho) dx^i$. Then, $\nabla_a \beta^a= \frac{1}{\sqrt{\gamma}} \frac{d}{d\rho} ( \sqrt{\gamma} \beta^{\rho})=0$, where $\gamma=\det \gamma_{ij}$, can be solved to get $\beta^{\rho}=c/\sqrt{\gamma}$ where $c$ is a constant. This constant is related to an invariant as follows: consider the scalar $i_{\beta} \star (m_1 \wedge \cdots\wedge m_{D-3})= c J$ where $\star$ is the Hodge dual with respect to the horizon metric $\gamma_{ab}$\footnote{We choose an orientation by setting $\epsilon_{\rho 1 \cdots D-3}=+\sqrt{\gamma}$.} and $J$ is the Jacobian of the coordinate transformation $\phi^i \to x^i$ (which is a non-zero constant). Therefore, if one of the rotational Killing fields on the horizon vanishes somewhere, we must have $c=0$ and hence $\beta_{\rho}\equiv 0$. This is indeed the case for all the topologies we are interested in (in fact $c \neq 0$ only in the $T^3$ case).
Thus,
exploiting the global representation for the one-form $h$ derived
above (\ref{globalhrep}), in the coordinates $(\rho,x^i)$ we have \be
 h= \Gamma^{-1} k_i(\rho) dx^i -\frac{\Gamma'}{\Gamma} d\rho
\ee
where in general we write $f' =df/d\rho$, and we have defined the functions $k_i(\rho)= \Gamma\beta_i$. Notice that the functions $k_i \equiv \Gamma h \cdot (\partial /\partial x^i)$ are in fact globally defined.

It is then convenient to introduce a new radial
coordinate (for the full near-horizon geometry) by $r \to \Gamma(\rho) r$ and define the function $A
\equiv \Gamma^2F-k^ik_i$, where $k^i \equiv \gamma^{ij}k_j$. One of
the main results found in~\cite{KLR} is that the vanishing of the
$\rho i$ and $\rho v$ components of the Ricci tensor of the full
near-horizon geometry in these new $(v,r,\rho,i)$ coordinates
implies that $k^i$ are constants and that $A =A_0 \Gamma$ for some
constant $A_0$. Then the near-horizon geometry written in the new
$r$ coordinate simplifies to \be \label{canNH} ds^2 = \Gamma(\rho)[
A_0 r^2 dv^2 +2dvdr]+ d\rho^2 + \gamma_{ij}(\rho)( dx^i+ k^i
rdv)(dx^j+k^jrdv) \; .\ee This form of the near-horizon geometry
makes an $SO(2,1)$ isometry group manifest~\cite{KLR}. We will take
this result as the starting point of our analysis and find all
vacuum geometries of this form with compact $\mathcal{H}$. For
completeness though, we will first show how this result is derived
from the Einstein equations written in terms of the near-horizon
data $(F,h_a,\gamma_{ab})$ defined purely on $\mathcal{H}$, i.e.
equations (\ref{Ric}) and (\ref{F}). First observe
that the $\rho i$ component of (\ref{Ric}) is $R_{\rho
i}=-\frac{1}{2} \Gamma^{-1}\gamma_{ij}(k^j)'$ and since $R_{\rho
i}=0$ for a metric of the form (\ref{hor}) this implies $k^i$ are
constants. Now, the $\rho$ component of (\ref{gradF}) is
$F'=-\frac{\Gamma'}{\Gamma}F+\Gamma^{-1}k^i(\Gamma^{-1}k_i)'$ which
when written in terms of the function $A$ defined above is
equivalent to $A'+\frac{\Gamma'}{\Gamma} A+ (k^i)'k_i=0$. From this
and the constancy of $k^i$ it follows that $A=A_0\Gamma$ for some
constant $A_0$ and the result is established\footnote{Since the functions $A$ and $\Gamma$ are globally defined the constant $A_0$ must be the same in every coordinate patch and thus this expression is valid globally.}.

Note that we will be only interested in non-static\footnote{A static
near-horizon geometry is defined by $V\wedge dV=0$, see~\cite{KLR}.}
near-horizon geometries, as the static case has been analysed
previously~\cite{CRT} where it was found that the only solution is a
direct product of $R^{1,1}$ and a flat compact space for $\Lambda
=0$, or a direct product of $AdS_2$ and a negative curvature compact
Einstein space for $\Lambda<0$. If all the constants $k^i=0$ then
the above near-horizon geometry is static~\cite{KLR}. Therefore we
will assume that at least one of the constants $k^i \neq 0$ in this
paper\footnote{Note this does not imply the vector field $k$ is
non-vanishing everywhere -- indeed we will find examples where $k$
has isolated zeros (which occurs at the fixed points of (a
combination of) the $U(1)^{D-3}$ Killing fields).}.

Let us now consider the near-horizon equation (\ref{F}). Observe
that $F= (A_0 \Gamma+ k^ik_i)/\Gamma^2$ and therefore (\ref{F})
becomes \be \label{A0} A_0 + \frac{k^ik_i}{2\Gamma}
-\frac{1}{2}{\nabla}^2 \Gamma =\Lambda \Gamma \;.\ee  Integrating (\ref{A0}) over $\mathcal{H}$ shows that
\be A_0 = \frac{1}{\textrm{vol}[\mathcal{H} ]} \int_{\mathcal{H}}
\left( -\frac{k^ik_i}{2\Gamma} +\Lambda \Gamma \right) \leq 0 \ee
with equality if and only if $\Lambda =0$ and $k^i=0$. Therefore for
non-static near-horizon geometries $A_0<0$ and we will often set
$A_0=-C^2$ for some $C>0$.

Now let us turn to the equation for the Ricci tensor of the horizon
(\ref{Ric}). The non-zero components of the Ricci tensor of the
horizon metric (\ref{hor}) are given by
\begin{eqnarray}
\label{Ricij} R_{ij} &=& -\frac{1}{2}\gamma_{ij}''- \frac{\gamma'}{4\gamma} \gamma_{ij}'+\frac{1}{2}\gamma_{ik}' \gamma^{kl} \gamma_{lj}' = -\frac{1}{2} \nabla^2 \gamma_{ij}+\frac{1}{2}\gamma_{ik}' \gamma^{kl} \gamma_{lj}', \label{Rij}\\
\label{Ricpp} R_{\rho\rho} &=& -\frac{1}{2} (\log \gamma)''-
\frac{1}{4} \gamma^{lj} \gamma_{jk}'\gamma^{km}\gamma_{ml}'.
\end{eqnarray}
and note that for
a function $f(\rho)$ \be \nabla^2 f \equiv f''+
\frac{\gamma'}{2\gamma}f' \ee where $\gamma \equiv \det
\gamma_{ij}$.
Evaluating the RHS of (\ref{Ric}) gives \bea \label{sourcepp}
R_{\rho \rho} &=& \frac{\Gamma''}{\Gamma} - \frac{1}{2}
\frac{\Gamma'^2}{\Gamma^2}+\Lambda, \\ \label{sourceij} R_{ij} &=&
\frac{1}{2} \Gamma^{-2} k_i k_j+ \frac{1}{2}
\gamma_{ij}'\frac{\Gamma'}{\Gamma}+ \Lambda\gamma_{ij} \; .\eea
Now, observe that (\ref{Ricij}) implies: \be R_{ij}\gamma^{ij} =
-\frac{1}{2}(\log \gamma)'' - \frac{1}{4}{(\log \gamma)'}^{2} \ee
which using (\ref{sourceij}) implies \be \label{det} (\log \gamma)''
+ \frac{\Gamma'}{\Gamma} (\log \gamma)'+ \Gamma^{-2}
k^2+\frac{1}{2}{(\log \gamma)'}^2+2(D-3)\Lambda=0. \ee By
contracting (\ref{Ricij}) with $k^ik^j$ and using (\ref{sourceij})
one gets: \be \label{Rickk} (k^2)''+ \frac{\Gamma'}{\Gamma} (k^2)' -
k_i' \gamma^{ij} k_j' + \frac{1}{2} (\log\gamma)'(k^{2})' + 2\Lambda
k^2+ \Gamma^{-2} (k^2)^2 =0 \ .\ee

To integrate the above equations it proves useful to use the
globally defined function $\sigma$ introduced at the beginning of this section as a coordinate instead of $\rho$. Note that in $\rho$ coordinates it is given by $\sigma' =\sqrt{\det\gamma_{ij}}$. Observe that
the volume form of $\mathcal{H}$ is then given simply by  $\epsilon_{D-3}= d\sigma \wedge dx^1 \wedge \cdots \wedge
dx^{D-3}$ (choosing
an orientation). Recall that $\sigma$ is a globally defined function and
$d\sigma$ is non-zero everywhere except where $\gamma_{ij}$
degenerates. Indeed, $\sigma$ cannot be constant as otherwise $\sigma'=0$ and thus $\det\gamma_{ij}=0$ everywhere.  Therefore it is legitimate to use $\sigma$ as a
coordinate everywhere except at these degeneration points (which occur at fixed points of $m_i$). Therefore any expression we derive in this coordinate will be valid globally provided we can show that is also smooth where $d\sigma=0$.

We will now derive some general results valid in both $D=4,5$.
Substituting into equation (\ref{A0}) implies \be \label{keq}
\frac{k^ik_i}{2\Gamma^2}= \frac{\Gamma' \sigma''}{2\Gamma
\sigma'}+\frac{C^2}{\Gamma}+ \frac{\Gamma''}{2\Gamma} +\Lambda  \ee
and equation (\ref{det}) gives \be \sigma''' +\frac{\Gamma'
\sigma''}{\Gamma} + \sigma'
\left(\frac{k^ik_i}{2\Gamma^2}+(D-3)\Lambda \right)=0 \; .\ee
Eliminating $k^ik_i$ between these two equations leads to \be
\label{sigmaeq} \sigma'''+ \frac{3 \Gamma'}{2\Gamma} \sigma'' +
\left( \frac{C^2}{\Gamma}+ \frac{\Gamma''}{2\Gamma} +(D-2)\Lambda
\right) \sigma'=0 \; .\ee This equation may actually be solved by
noting the identity \be \sigma'''+ \frac{3 \Gamma'}{2\Gamma}
\sigma'' + \left( \frac{C^2}{\Gamma}+ \frac{\Gamma''}{2\Gamma}
+(D-2)\Lambda \right) \sigma' \equiv \sigma' \left[
\frac{1}{2\Gamma} \frac{d^2Q}{d\sigma^2}  + \frac{C^2}{\Gamma}
+(D-2)\Lambda \right] \ee where we have defined $Q(\sigma) \equiv
\sigma'^2 \Gamma$. Therefore we deduce that \be \label{ddQ} \ddot{Q}
+ 2C^2 +2(D-2) \Lambda \Gamma =0 \; .\ee where in general we denote
$df/d\sigma=\dot{f}$. Observe that by working in the $\sigma$
coordinate the $d\rho^2$ part of the metric is given by \be
\label{gpp} d\rho^2= \frac{\Gamma }{Q} d\sigma^2 \; . \ee
Substituting $\sigma'^2=Q/\Gamma$ back into (\ref{keq}) gives \be
\label{kiki} k^ik_i = \Gamma \frac{d}{d\sigma} \left( \frac{Q \dot{\Gamma}}{\Gamma} \right)+2C^2 \Gamma+2\Lambda \Gamma^2 \; .\ee
Since we are assuming the constants $k^i \neq 0$, we can always
choose the coordinates $x^i$ such that $k^i\partial/\partial x^i =
\partial/\partial x^1$. This implies $k^ik_i = \gamma_{11}$ and
therefore we have determined this component of the metric in terms
of the functions $Q$ and $\Gamma$. In $D=4$, together with
(\ref{gpp}), this determines the whole metric on $\mathcal{H}$ in
terms of the two functions $Q$ and $\Gamma$.

Before closing this section let us derive a useful result based on
global considerations. First notice that the norm of the one form
$d\sigma$ is given by \be (d\sigma)^2 = \frac{Q}{\Gamma} \; \ee
which implies $Q \geq 0$. Now since $\sigma$ is a globally defined
function on a closed manifold $\mathcal{H}$ it must have a distinct
minimum (say $\sigma_1$) and maximum (say $\sigma_2$) so $\sigma_1
\leq \sigma \leq \sigma_2$  and $\sigma_1< \sigma_2$ (note $\sigma$
cannot be a constant). Therefore $d\sigma$ must vanish at these two
distinct points on $\mathcal{H}$, and as argued earlier there are in
fact only two points where $d\sigma=0$ (they correspond to the fixed points of $m_i$). This
implies that the function $Q \geq 0$ with equality if and only if
$\sigma=\sigma_1$ or $\sigma=\sigma_2$. We deduce that any function
we construct from $\sigma$-derivatives of globally defined functions
can only fail to be defined at the points
$\sigma=\sigma_1,\sigma_2$. Our subsequent analysis will be mostly
local (integration of ODEs wrt $\sigma$), although there are steps
where we need to use the fact that certain functions are globally
defined. In the Appendix we introduce a (globally defined) vector
field which allows us to prove that these functions are globally
defined.

\section{Four dimensions}
In four dimensions the metric on the horizon is particularly simple
\be
\gamma_{ab} dx^a dx^b = d\rho^2 + \gamma(\rho) dx^2
\ee
where we write $x^1=x$ and note that $\gamma_{11}= \gamma$ in this case. Equation (\ref{kiki}) therefore gives an expression for $\gamma$ which, noting that $\gamma= \sigma'^2=Q/\Gamma$, can be written as
\be
\label{Qeq} Q = \dot{Q} \dot{\Gamma} \Gamma - \dot{\Gamma}^2
Q +Q \Gamma \ddot{\Gamma}+2C^2 \Gamma^2+2\Lambda \Gamma^3 \; .\ee
Now differentiate (\ref{Qeq}) with respect to $\sigma$. This gives an expression
involving $\ddot{Q}$ which can be eliminated using (\ref{ddQ})
leaving: \be \dot{Q} = Q\Gamma \frac{d^3\Gamma}{d\sigma^3}
+\ddot{\Gamma}(2\dot{Q}\Gamma-Q\dot{\Gamma})+2C^2\Gamma
\dot{\Gamma}+2\Lambda\Gamma^2 \dot{\Gamma} \; .\ee Now combine this with
(\ref{Qeq}) in such a way to eliminate the $C^2$ and $\Lambda$
terms to eventually get \be \label{dddGamma} Q\frac{d^3
\Gamma}{d\sigma^3} + \left( \dot{Q}-
\frac{\dot{\Gamma Q}}{\Gamma} \right)\left( 2\ddot{\Gamma}-
\frac{\dot{\Gamma}^2}{\Gamma}- \frac{1}{\Gamma} \right) =0 \; .\ee Now
define \be \mathcal{P} \equiv 2\ddot{\Gamma}- \frac{\dot{\Gamma}^2}{\Gamma}-
\frac{1}{\Gamma} \ee and note the identity \be \label{Pid}
2\frac{d^3 \Gamma}{d\sigma^3} \equiv \frac{\dot{\Gamma} \mathcal{P}}{\Gamma}+
\dot{\mathcal{P}} \;.\ee Eliminate the third order derivative terms between
(\ref{dddGamma}) and (\ref{Pid}) to get: \be \dot{\mathcal{P}} =\left(
\frac{\dot{\Gamma}}{\Gamma} - \frac{2 \dot{Q}}{Q} \right) \mathcal{P} \ee
which integrates to \be \label{alphaeq} \frac{Q^2 \mathcal{P}}{\Gamma} =\alpha \ee where
$\alpha$ is some constant.

As discussed earlier, based on global analysis $Q$ must vanish at two distinct points which from (\ref{alphaeq}) would seem to say one must have $\alpha=0$. Indeed, in the Appendix we prove that $Q^2\mathcal{P}$ is a globally defined function which vanishes at the zeros of $Q$ and therefore one must have $\alpha=0$ for a compact $\mathcal{H}$. From (\ref{alphaeq}) we see that therefore we must have  $\mathcal{P}=0$, and this equation can be solved noting the identity \be \frac{\mathcal{P} \dot{\Gamma} }{\Gamma} \equiv \frac{d}{d\sigma}
\left( \frac{ \dot{\Gamma}^2+1}{\Gamma}\right) \;, \ee  which implies \be \dot{\Gamma}^2+1= \beta \Gamma
\ee where $\beta>0$ is a constant. There are two solutions to this
equation: either \be \label{Gamma4d} \Gamma = \beta^{-1}+ \frac{\beta
(\sigma-\sigma_0)^2}{4} \ee where $\sigma_0$ is a constant, or simply
$\Gamma=\beta^{-1}$. This latter solution implies $\Gamma$ and $Q$ are both constants -- this is incompatible with having a compact $\mathcal{H}$ and therefore we discount it. Therefore $\Gamma$ must be given by (\ref{Gamma4d}) and since by definition $\sigma$ is only defined up to an additive constant, without loss of generality we will set $\sigma_{0}=0$.  We can now integrate easily for $Q$ using (\ref{ddQ}) to find:
\begin{equation}
Q = -\frac{\beta \Lambda}{12}\sigma^4 - (C^2 + 2\Lambda\beta^{-1})\sigma^2 + c_1\sigma + c_2 \; .
\end{equation}  Now plugging back into equation (\ref{Qeq}) implies
\begin{equation}
c_2 = 4\beta^{-3}(C^2\beta + \Lambda) \; .
\end{equation}
The rest of the near-horizon equations are now satisfied without further constraint.

To summarise, so far we have shown that the near-horizon geometry is
given by \be ds^2= \Gamma[ -C^2r^2 dv^2+2dvdr]
+\frac{\Gamma}{Q}d\sigma^2 + \frac{Q}{\Gamma} (dx+rdv)^2 \ee where
\be \Gamma=\beta^{-1}+ \frac{\beta \sigma^2}{4}, \qquad Q =
-\frac{\beta \Lambda}{12}\sigma^4 - (C^2 +
2\Lambda\beta^{-1})\sigma^2 + c_1\sigma + 4\beta^{-3}(C^2\beta +
\Lambda) \ee and $C>0$, $\beta>0$ and $c_1$ are constants. Observe
that the near-horizon geometry has the following scaling freedom \be
\label{scale4} C^2 \to K C^2, \qquad \beta \to K^{-1} \beta, \qquad
c_1 \to K^2 c_1 \qquad \sigma \to K \sigma, \qquad x \to K^{-1} x,
\qquad v \to K^{-1} v \ee for constant $K>0$, which allows one to
fix one of the parameters (or a combination of them) to any desired
value.

 Although we have used some global information in our derivation, we need to complete the global analysis of this solution to determine the most general regular near-horizon geometry with compact horizon sections.

\subsection{Global analysis}
Consider the metric on $\mathcal{H}$: \be \gamma_{ab}dx^a dx^b =
\frac{\Gamma}{Q} d\sigma^2 + \frac{Q}{\Gamma} dx^2 \; . \ee As
discussed earlier compactness of $\mathcal{H}$ requires $\sigma_1
\leq \sigma \leq \sigma_2$ and $Q \geq 0$ with equality occurring at
$\sigma_1,\sigma_2$ only. It follows that $\dot{Q}(\sigma_1)>0$ and
$\dot{Q}(\sigma_2)<0$. The Killing vector $\partial / \partial x$
must vanish at the endpoints. The horizon metric therefore is
non-degenerate everywhere except at $\sigma=\sigma_1,\sigma_2$ where
in general one has conical singularities. Simultaneous removal of
the conical singularities at $\sigma_1$ and $\sigma_2$ is equivalent
to \be \label{balance}\frac{\dot{Q}(\sigma_1)}{\Gamma(\sigma_1)} =
-\frac{\dot{Q}(\sigma_2)}{\Gamma(\sigma_2)} \; . \ee If this
condition is satisfied we have a regular metric with $\partial
/\partial x$ vanishing at the endpoints $\sigma=\sigma_1,\sigma_2$
and therefore $\mathcal{H}$ has $S^2$ topology as expected.

Let us first consider $\Lambda=0$ so $Q(\sigma)=-C^2
\sigma^2+c_1\sigma+c_2= C^2(\sigma-\sigma_1)(\sigma_2-\sigma)$. It
follows that $\dot{Q}(\sigma_1)=-\dot{Q}(\sigma_2)$ and therefore
using the condition for the absence of conical singularities
(\ref{balance}) we have $\Gamma(\sigma_1)=\Gamma(\sigma_2)$. Since
the roots must be distinct, using the form of $\Gamma$ we see that
$\sigma_1=-\sigma_2 \neq 0$. This implies $c_1=0$ and from the
expression for $c_2$ we get $\sigma_1=-2\beta^{-1}$ so $Q=C^2(
4\beta^{-2} -\sigma^2)$. Define a new coordinate $\phi= C^2 x$, a
parameter $a=\frac{1}{C \sqrt{\beta}}$ and rescale $\sigma \to
2\sigma/ \beta$. The horizon metric then becomes \be \gamma_{ab}
dx^a dx^b = a^2 \left(\frac{1+\sigma^2}{1-\sigma^2} \right)d\sigma^2
+ 4a^2 \left(\frac{1-\sigma^2}{1+\sigma^2}\right) d\phi^2 \ee and
regularity implies $\phi$ to be $2\pi$ periodic. This is an
inhomogeneous metric on $S^2$ with $\partial / \partial \phi$
vanishing at $\sigma=\pm 1$. The full near-horizon geometry, upon
rescaling $v \to \beta v/2$, is now given by: \be ds^2
=\frac{1+\sigma^2}{2}\left[ - \frac{r^2}{2a^2} dv^2 +2dvdr \right]+
a^2 \left(\frac{1+\sigma^2}{1-\sigma^2} \right)d\sigma^2 + 4a^2
\left(\frac{1-\sigma^2}{1+\sigma^2}\right) \left(
d\phi+\frac{r}{2a^2} dv \right)^2 \; . \ee This coincides exactly
with the near-horizon geometry of extremal Kerr as given
in~\cite{KLR} upon the change of variables $\sigma=\cos\theta$. This
proves that: \\

\noindent {\it The only 4d Ricci flat axisymmetric near-horizon geometry
with a non-toroidal horizon section is that of the extremal Kerr black hole}. \\

Now consider the $\Lambda <0$ case and set $\Lambda=-3g^2$. We have
argued that $Q$ must have distinct roots $\sigma_1<\sigma_2$ and be
positive in the interval in between these roots. Therefore, since $Q$ is a
quartic with a positive $\sigma^4$ coefficient, it must have four real roots and further they must
be all distinct (for compactness), such that
$\sigma_0<\sigma_1<\sigma_2<\sigma_3$. Therefore
\begin{equation}
Q = \frac{\beta
g^2}{4}(\sigma-\sigma_0)(\sigma-\sigma_1)(\sigma-\sigma_2)(\sigma-\sigma_3)
\end{equation} and due to the absence of a cubic term in $Q$ we must
have \be \sigma_1+\sigma_2+\sigma_3+\sigma_0=0 \; .
\label{sumroot}\ee The condition for the absence of conical
singularities (\ref{balance}) becomes \be \label{conic}
\frac{(\sigma_2-\sigma_0)(\sigma_3-\sigma_2)}{(\sigma_1-\sigma_0)(\sigma_3-\sigma_1)}
= \frac{\Gamma(\sigma_2)}{\Gamma(\sigma_1)} \; .\ee Now, we prove
that this implies $\Gamma(\sigma_1)=\Gamma(\sigma_2)$. To do so,
first assume $\Gamma(\sigma_2)>\Gamma(\sigma_1)$. This implies that
the LHS of (\ref{conic}) is greater than one and in turn this
implies, using (\ref{sumroot}), that $\sigma_2+\sigma_1<0$ and thus
$\sigma_2^2<\sigma_1^2$. It follows that
$\Gamma(\sigma_2)<\Gamma(\sigma_1)$ in contradiction to our
assumption. Similarly assuming $\Gamma(\sigma_2)<\Gamma(\sigma_1)$
implies $\sigma_1+\sigma_2>0$ and hence $\sigma_2^2>\sigma_1^2$
providing another contradiction. We conclude that
$\Gamma(\sigma_1)=\Gamma(\sigma_2)$ and hence $\sigma_2=-\sigma_1
\neq 0$. From (\ref{sumroot}) it follows that $\sigma_3=-\sigma_0$
and therefore $Q$ is an even function of $\sigma$, i.e. $c_1=0$.

Now we will show that the\footnote{In fact the $c_1 \neq 0$ solution is the near-horizon limit of Kerr-AdS$_4$-NUT whose horizon suffers from conical singularities.} $c_1=0$ solution is the near-horizon limit
of Kerr-AdS$_4$. Comparing coefficients of $Q$ gives \bea
\sigma_2^2+\sigma_3^2=\frac{4 C^2}{g^2\beta } -\frac{24}{\beta^2},
\\ \sigma_2^2\sigma_3^2= \frac{16C^2}{g^2
\beta^3}-\frac{48}{\beta^4} \; .\eea These two equations are
equivalent to \bea (\beta \sigma_2)^2(\beta \sigma_3)^2-4(\beta
\sigma_2)^2-4(\beta \sigma_3)^2=48, \label{cond1}\\
\label{cond2} (\beta \sigma_2)^2(\beta \sigma_3)^2-2(\beta
\sigma_2)^2-2(\beta \sigma_3)^2= \frac{8C^2 \beta}{g^2} \; .\eea Now
define two positive constants $a,r_+$ by \be a \equiv
\frac{\sigma_2}{g \sigma_3}, \qquad r_+ \equiv \frac{2}{g \beta
\sigma_3} \ee and so it follows $ag <1$. Note that the parameters $a$
and $r_+$ are actually invariant under the scale transformation
(\ref{scale4}). Using these definitions to eliminate
$\sigma_2,\sigma_3$ from (\ref{cond1}) implies $g^2r_+^2<1$ and \be a^2 =
\frac{r_+^2(1+3g^2r_+^2)}{1-g^2 r_+^2} \label{extr} \; .\ee
Next, eliminate $\sigma_2,\sigma_3$ in (\ref{cond2})
and then use the expression for $a$ (\ref{extr}) to get \be
\label{betaC2} \beta C^2 =
\frac{1+6g^2r_+^2-3g^4r_+^4}{r_+^2(1-g^2r_+^2)} =
\frac{1+a^2g^2+6g^2r_+^2}{r_+^2} \; .\ee Next use the scale
invariance (\ref{scale4}) of the near-horizon geometry to set \be
C^2=\frac{1+a^2g^2+6g^2r_+^2}{\Xi(r_+^2+a^2)} \ee where we define
$\Xi \equiv 1-a^2g^2$. Using this choice of $C^2$ (\ref{betaC2})
implies \be \beta = \frac{\Xi(r_+^2+a^2)}{r_+^2} \; .\ee Plugging
this into the definition of $r_+$ gives \be \sigma_3=
\frac{2r_+}{g\Xi (r_+^2+a^2)} \ee and then from the definition of
$a$ it follows that \be \sigma_2=\frac{2r_+a}{\Xi (r_+^2+a^2)} \;
.\ee Finally change coordinates from $(\sigma, x)$ to
$(\theta,\phi)$ defined by \be \phi= \frac{2ar_+ x}{(r_+^2+a^2)^2},
\qquad \cos\theta = \frac{\sigma}{\sigma_2} \ee so $ 0 \leq \theta
\leq \pi$ provides a unique parametrization of the interval. This
gives \be Q= \frac{4r_+^2a^2 \sin^2\theta
\Delta_{\theta}}{\Xi^3(r_+^2+a^2)^3}, \qquad \Gamma=
\frac{\rho_+^2}{\Xi(r_+^2+a^2)} \ee where $\Delta_{\theta}=1-a^2g^2
\cos^2\theta$ and $\rho_+^2= r_+^2 +a^2 \cos^2\theta$. It follows
that \be \frac{\Gamma d\sigma^2}{Q}+ \frac{Q}{\Gamma} dx^2=
\frac{\rho_+^2 d\theta^2}{\Delta_{\theta}}+\frac{\sin^2\theta
\Delta_{\theta} (r_+^2+a^2)^2}{\rho_+^2 \Xi^2} d\phi^2 \; \ee
and it is easy to see that absence of conical singularities implies $\phi \sim \phi +2\pi$.
Inspecting the Appendix we see that this is exactly the horizon
geometry of
Kerr-AdS$_4$ and the rest of the near-horizon data $\Gamma,k^{\phi}$ also agrees. Therefore we have proved that:
\\ \\ \noindent { \it The only four dimensional axisymmetric near-horizon geometry with a non-toroidal horizon section which satisfies $R_{\mu\nu}=\Lambda g_{\mu\nu}$, with $\Lambda<0$, is the
near-horizon limit of Kerr-AdS$_4$.} \\

This completes the proof of Theorem 1 stated in the Introduction.

\section{Five dimensions}

\subsection{Near-horizon equations}
In five dimensions it is useful to re-write the horizon metric as
\be \label{5d} \gamma_{ab} dx^a dx^b = d\rho^2+ \gamma_{11}(\rho) ( dx^1
+\omega(\rho) dx^2)^2 + \frac{\gamma(\rho)}{\gamma_{11}(\rho)}
(dx^2)^2 \ee where we define $\omega(\rho) \equiv
\gamma_{12}/\gamma_{11}$ and recall $\gamma = \det \gamma_{ij}$. We
have already determined $k^ik_i=\gamma_{11}$ in terms of $\Gamma,Q$
(\ref{kiki}). Since we also know $\gamma=\sigma'^2=Q/\Gamma$ we need
to determine only one other component of $\gamma_{ij}$, say
$\gamma_{12}$ or equivalently $\omega$.

Consider (\ref{Rickk}), which since we have chosen $k=\partial/\partial x^1$ is equivalent to the $R_{11}$ equation. To simplify this equation we will need the identity
\be
k_i'\gamma^{ij} k_j' \equiv \frac{(\gamma_{11}')^2}{\gamma_{11}} + \frac{\gamma_{11}^3}{\gamma}\left[\left(\frac{\gamma_{12}}{\gamma_{11}}\right)'\right]^2 \; ,
\ee
substitute for $\gamma=\sigma'^2=Q/\Gamma$, convert all $\rho$ derivatives to $\sigma$ derivatives and note the fact $\sigma''=\frac{d}{d\sigma} \left(\frac{Q}{2\Gamma} \right)$. The result is that (\ref{Rickk}) becomes
\be
\label{omegadot1}
\gamma_{11}^2 \dot{\omega}^2= \frac{1}{\Gamma} \frac{d}{d\sigma}\left( \frac{Q \dot{\gamma}_{11}}{\gamma_{11}} \right) +2\Lambda+ \frac{\gamma_{11}}{\Gamma^2} \;.
\ee

Now consider the $\rho\rho$ component of (\ref{Ric}) which is given by equating (\ref{Ricpp}) and (\ref{sourcepp}). To evaluate (\ref{Ricpp}) it proves useful to note the identity
\begin{eqnarray}
\label{gamma4}
\gamma^{lj}\gamma'_{jk}\gamma^{km}\gamma'_{ml} & \equiv& \frac{(\gamma_{11}')^2}{ \gamma_{11}^2} +\left( \frac{\gamma_{11}'}{\gamma_{11}} -\frac{\gamma'}{\gamma} \right)^2 + \frac{2\gamma_{11}^2}{\gamma}\left[\left(\frac{\gamma_{12}}{\gamma_{11}}\right)'\right]^2 \\ &=&  \frac{Q(\dot{\gamma_{11}})^2}{\Gamma\gamma_{11}^2} + \frac{Q}{\Gamma}\left(\frac{\dot{\gamma_{11}}}{\gamma_{11}} - \frac{\Gamma}{Q}\frac{d}{d\sigma}\left(\frac{Q}{\Gamma}\right)\right)^2 + 2\gamma_{11}^2\dot{\omega}^2 \nonumber
\end{eqnarray}
where in the second line we have converted to $\sigma$ derivatives. The other term in (\ref{Ricpp}) is given by $\log \gamma''$, which using $\gamma=\sigma'^2$ contains a $\sigma'''$ and we eliminate this using (\ref{sigmaeq}).  After some calculation the $\rho\rho$ equation simplifies to
\be
\label{omegadot2}
\gamma_{11}^2 \dot{\omega}^2= \frac{2C^2}{\Gamma}+4\Lambda- \frac{Q \ddot{\Gamma}}{\Gamma^2}+ \frac{\dot{Q} \dot{\Gamma}}{\Gamma^2}- \frac{Q \dot{\gamma}_{11}^2}{\Gamma \gamma_{11}^2} + \frac{\dot{\gamma_{11}}}{\gamma_{11}}\frac{d}{d\sigma}\left( \frac{Q}{\Gamma} \right) \; .
\ee

Equating (\ref{omegadot1}) and (\ref{omegadot2}), using (\ref{kiki}) to write the $\gamma_{11}/\Gamma^2$ term in (\ref{omegadot1}), leads to
\be
\label{gamma11ode}
\frac{d}{d\sigma} (\Gamma \dot{\gamma}_{11})+ \left(2\ddot{\Gamma}-\frac{\dot{\Gamma}^2}{\Gamma} \right)\gamma_{11}=0 \; .
\ee
Differentiating (\ref{kiki}) with respect to $\sigma$ gives
\be
\label{gamma11dot}
\dot{\gamma}_{11} = Q\frac{d^3\Gamma}{d\sigma^3} + \left( 2\ddot{\Gamma}-\frac{\dot{\Gamma}^2}{\Gamma} \right)\left( \dot{Q}- \frac{Q \dot{\Gamma}}{\Gamma} \right)  -2\Lambda \Gamma \dot{\Gamma}
\ee
where (\ref{ddQ}) has been used to eliminate $\ddot{Q}$. Substituting (\ref{gamma11dot}) and (\ref{kiki}) into (\ref{gamma11ode}), again using (\ref{ddQ}) to eliminate $\ddot{Q}$, leads to the remarkably simple equation
\be
Q\frac{d^4 \Gamma }{d\sigma^4}+ \left( 3 \dot{Q} - \frac{\dot{\Gamma}Q}{\Gamma} \right) \frac{d^3\Gamma}{d\sigma^3} -10 \Lambda\Gamma\ddot{\Gamma}
 =0
\ee
which can be written more compactly as
\be
\label{Gammaode}
\frac{d}{d\sigma} \left( \frac{Q^3}{\Gamma} \frac{d^3 \Gamma}{d\sigma^3} \right) - 10 \Lambda Q^2 \ddot{\Gamma}=0 \; .
\ee

We must now examine the remaining components of the near-horizon equations, i.e. the $x^1x^2$ components of (\ref{Ric}). One can check
that the $12$ component of (\ref{Ricij}) is
\begin{eqnarray}
R_{12} &=& -\frac{\gamma_{11}''\omega}{2} - \gamma_{11}'\omega' -\frac{\gamma_{11}\omega''}{2} + \frac{\gamma_{11}\gamma'\omega'}{4\gamma} + \frac{\gamma_{11}^3\omega \omega'^2}{2\gamma} + \frac{\gamma_{11}'^2\omega}{2\gamma_{11}} - \frac{\gamma_{11}'\gamma'\omega}{4\gamma} \nonumber \\
& = & -\frac{Q\ddot\gamma_{11} \omega}{2\Gamma} -
\frac{d}{d\sigma}\left(\frac{Q}{\Gamma}\right)\frac{\dot\gamma_{11}\omega}{2}
+ \frac{\gamma_{11}^3\omega \dot\omega^2}{2} -
\frac{Q\dot\gamma_{11}\dot\omega}{\Gamma} -
\frac{Q\gamma_{11}\ddot\omega}{2\Gamma} +
\frac{Q\dot\gamma_{11}^2\omega}{2\Gamma \gamma_{11}} \label{R12}
\end{eqnarray}
and (\ref{sourceij}) requires that
\begin{equation}\label{R12source}
R_{12} = \frac{\gamma_{11}^2\omega}{2\Gamma^2} + \frac{Q\dot\Gamma}{2\Gamma^2}\left[\dot\omega \gamma_{11} + \omega \dot\gamma_{11}\right] +\Lambda \gamma_{11} \omega \; .
\end{equation}
Eliminating the $\dot{\omega}^2$ term in (\ref{R12}) using (\ref{omegadot1}) leads to many cancellations and the $x^1x^2$ component of (\ref{Ric}) becomes simply
\be
\gamma_{11}\ddot{\omega}+2\dot{\gamma}_{11}\dot{\omega}+\frac{\dot{\Gamma}}{\Gamma} \gamma_{11} \dot{\omega}=0
\ee
which integrates to
\be
\label{intcond}
\dot{\omega} =\frac{k}{\gamma_{11}^2\Gamma}
\ee
where $k$ is a constant. In fact (\ref{intcond}) is automatically satisfied as a consequence of the other components of the near-horizon equations. Indeed, using (\ref{omegadot2}) and (\ref{kiki}) one can check that $\frac{d}{d\sigma} (\gamma_{11}^4\Gamma^2 \dot{\omega}^2)=0$ as a consequence of (\ref{ddQ}) and (\ref{Gammaode}).

In fact the above equations exhibit certain scaling symmetries which
translate to scaling symmetries of the full near-horizon geometry.
It is important to keep track of these when it comes to counting the
parameters of a solution. The two ODEs (\ref{ddQ}) and
(\ref{Gammaode}) possess the following two symmetries: \bea
\label{S1}
&&\mathcal{S}_1: \qquad Q \to K^3Q, \qquad \Gamma \to K \Gamma, \qquad C^2 \to K C^2, \qquad \sigma \to K\sigma \\
&&\mathcal{S}_2: \qquad Q \to L^2 Q, \qquad \sigma \to L\sigma
\label{S2}\eea for constant $K>0$ and constant $L$ (of either sign).
It follows that \bea
&&\mathcal{S}_1 : \qquad \gamma_{11} \to K^2 \gamma_{11}, \qquad x^1 \to K^{-1}x^1, \qquad v \to K^{-1}v \\
&&\mathcal{S}_2: \qquad \gamma_{12} \to L \gamma_{12}, \qquad  x^2 \to L^{-1} x^2
\eea
provide scaling symmetries of the full near-horizon geometry. Observe that these scalings can be combined, e.g. $\mathcal{S}_2^{-1}\mathcal{S}_1$ (with $K=L$) generates the near-horizon symmetry $Q\to KQ, \; \Gamma\to K\Gamma, \; C^2 \to KC^2, \; x^1 \to K^{-1}x^2 ,\; x^2 \to K x^2, \; v \to K^{-1}v$.

\paragraph{Summary} We have shown that the functions $\Gamma(\sigma)$ and $Q(\sigma)$ satisfy the coupled ODEs (\ref{ddQ}) and (\ref{Gammaode}). Further, given a solution to these ODEs $(\Gamma(\sigma),Q(\sigma))$, a near-horizon geometry satisfying the vacuum Einstein equations $R_{\mu\nu}=\Lambda g_{\mu\nu}$ can be constructed as follows. Firstly $\gamma_{11}$ is determined from (\ref{kiki}); next $\omega=\gamma_{12}/\gamma_{11}$ can be got up to quadratures from either (\ref{omegadot1}) or (\ref{omegadot2}); finally note (\ref{gpp}) gives $\gamma_{\sigma\sigma}$. This determines the horizon metric (\ref{5d}) in the coordinates $(\sigma,x^1,x^2)$. Recalling that we chose a gauge where $k^i =\delta^i_1$, one can write down the full near-horizon geometry from (\ref{canNH}).

\subsection{A class of near-horizon geometries with $S^3$ horizons}
Observe that one set of solutions to (\ref{Gammaode}) is given by:
\be \Gamma =a_1\sigma+a_0 \ee where $a_1,a_0$ are constants. Then,
(\ref{ddQ}) implies: \be \label{QMP}Q= -\Lambda a_1 \sigma^3 -
(C^2+3\Lambda a_0)\sigma^2 +c_1 \sigma +c_2 \ee where $c_1,c_2$ are
integration constants. The analysis naturally splits into two, depending
on whether $a_1$ vanishes or not\footnote{We could consider the two cases simultaneously, however for clarity we have chosen not to.}.

\subsubsection{Homogeneous horizon}
 First, suppose $a_1=0$ and so $\Gamma$ is a constant. Then, the
equation for $k^ik_i$ (\ref{kiki}) gives \be \gamma_{11}= 2C^2
\Gamma+ 2\Gamma^2 \Lambda \ee which is a constant and thus
$C^2+\Lambda\Gamma>0$. Equation (\ref{omegadot1}) gives \be
\dot{\omega}^2 = \frac{ (C^2+2\Lambda \Gamma)}{2\Gamma^3
(C^2+\Lambda \Gamma)^2}\ee which is also a constant and implies
$C^2+2\Lambda\Gamma \geq 0$. Therefore \be \omega =\pm \left({\frac{
(C^2+2\Lambda \Gamma)}{2\Gamma^3 (C^2+\Lambda \Gamma)^2}}
\right)^{\frac{1}{2}} \sigma + c_3 \ee where $c_3$ is an integration
constant. We may set $c_3=0$ using the coordinate freedom of the
$x^1 \to x^1+ \const x^2$ which we will now assume we have done.
Note that $Q=-(C^2+3\Lambda \Gamma) \sigma^2+c_1 \sigma+c_2$ and
since $\sigma$ is only defined up to an additive constant, without
loss of generality we may translate $\sigma$ in order to set
$c_1=0$. This implies $Q=c_2-(C^2+3\Lambda \Gamma) \sigma^2$. Recall
that in order to have a compact horizon one needs $\sigma_1\leq
\sigma \leq \sigma_2$ with $Q \geq 0$ in this interval and vanishing
only at the endpoints. It is easy to see this implies
$C^2+3\Lambda\Gamma>0$ (which is automatic when $\Lambda=0$). It now
follows that $c_2>0$ and $\sigma_2=-\sigma_1=\sqrt{c_2(C^2+3\Lambda
\Gamma)^{-1}}$. We now define new coordinates $(\theta,\psi,\phi)$
as follows: \be \cos\theta = \frac{\sigma}{\sigma_2}, \qquad \phi =\pm
x^2 \sqrt{ \frac{ c_2 (C^2+3\Lambda\Gamma)}{2\Gamma^3 (C^2+\Lambda
\Gamma)}}, \qquad \psi = x^1(C^2 +3\Lambda \Gamma) \sqrt{
\frac{C^2+\Lambda\Gamma}{C^2+2\Lambda \Gamma}}\ee so that $0\leq
\theta \leq \pi$ parameterizes the interval $\sigma_1 \leq \sigma
\leq \sigma_2$ uniquely and $Q=c_2 \sin^2\theta$. The near-horizon
data is then given by \bea
\label{NHdataGammaconst}\gamma_{ab}dx^adx^b &=&
\frac{2\Gamma(C^2+2\Lambda \Gamma)}{(C^2+3\Lambda\Gamma)^2}
(d\psi+\cos\theta d\phi)^2+ \frac{\Gamma}{C^2+3\Lambda
\Gamma}(d\theta^2+\sin^2\theta d\phi^2),
\\ k^{\psi} &=& (C^2 +3\Lambda \Gamma) \sqrt{
\frac{C^2+\Lambda\Gamma}{C^2+2\Lambda \Gamma}} \eea with $\Gamma$ a
constant. It is clear that regularity of the metric on $\mathcal{H}$
implies the usual restrictions $0 \leq \psi \leq 4\pi$ and $0 \leq
\phi \leq 2\pi$ resulting in a homogeneous metric on $S^3$ written
in Euler angles. This near-horizon geometry has the scaling symmetry
\be C^2 \to K C^2, \qquad \Gamma \to K \Gamma,\qquad v \to K^{-1}v
\ee where $K>0$ is a constant. This allows one to fix one (or a
combination) of the parameters $(C^2,\Gamma)$ of the above solution
and therefore it is a 1-parameter family. In fact, as we show in the
Appendix it is isometric to the near-horizon limit of the extremal
self-dual rotating $AdS_{5}$ black hole~\cite{HHT} (i.e. with $J_1=J_2$). In the case $\Lambda=0$ it turns out (as we also show in the Appendix) it is also isometric to the near-horizon limit of the $J=0$ extremal KK black hole~\cite{Rasheed}.

\subsubsection{Inhomogeneous horizon}
Now, suppose $a_1 \neq 0$. We are free to perform a translation in
$\sigma$ to set $a_0 = 0$, which without loss of generality we will
do. The equation for $k^ik_i$ (\ref{kiki}) gives: \be
\gamma_{11}=a_1\left( C^2\sigma- \frac{c_2}{\sigma} \right) \; . \ee
We can now solve for $\omega$ using~(\ref{omegadot2}). After some
calculation, equation (\ref{omegadot2}) gives
\begin{equation}
\dot \omega^2 = \frac{4\sigma^2 c_2(\Lambda a_1c_2 - c_1C^2)}{a_1^3(C^2 \sigma^2 - c_2)^4}
\end{equation}
and therefore the parameters must satisfy the inequality: \be
\label{ineq1} c_2(\Lambda a_1c_2 - c_1C^2) \geq 0 \; .\ee
Integrating one gets
\begin{equation}
\omega = \pm \frac{\sqrt{a_1^{-3}c_2(\Lambda a_1c_2 - c_1C^2)}}{C^2
(C^2 \sigma^2 - c_2)} + c_3
\end{equation} where $c_3$ is a constant. Collecting the above results
the horizon metric is: \be \gamma_{ab}dx^a dx^b = \frac{a_1 \sigma
d\sigma^2}{Q(\sigma)} + a_1\left( C^2\sigma- \frac{c_2}{\sigma}
\right) \left( dx^1 + \frac{\sqrt{a_1^{-3}c_2(\Lambda a_1c_2 -
c_1C^2)}}{C^2 (C^2 \sigma^2 - c_2)} \; dx^2 \right)^2 +
\frac{Q(\sigma) (dx^2)^2}{a_1^2 \left( C^2 \sigma^2
-c_2 \right)} \ee where by shifting $x^1 \to x^1
+\const x^2$ we have eliminated the constant $c_3$, used the freedom $x^2 \to \pm x^2$ to arrange $\omega>0$, and \be Q=
-\Lambda a_1 \sigma^3 -C^2 \sigma^2+c_1 \sigma+c_2 \; .\ee This
near-horizon metric has two independent scaling symmetries (corresponding to $\mathcal{S}_1$ and $\mathcal{S}_2$): \be \label{scaling1} C^2
\to K C^2, \qquad c_1 \to K^2 c_1, \qquad c_2 \to K^3 c_2, \qquad
\sigma \to K \sigma, \qquad x^1 \to K^{-1} x^1, \qquad v \to K^{-1}
v \ee where $K>0$ is constant, and \be a_1 \to L^{-1} a_1, \qquad c_1 \to Lc_1, \qquad c_2
\to L^2 c_2, \qquad \sigma \to L \sigma, \qquad x^2 \to L^{-1}x^2  \label{scaling2}
\ee where $L$ is constant (which can be either sign). These allow one to fix two (or two
combinations) of the parameters $(C^2,a_1,c_1,c_2)$ and thus this
solution is a 2-parameter family.

\subsubsection{Global analysis of inhomogeneous horizon}
We now turn to a global analysis of the $a_1 \neq 0$ solution just
derived. First we will use the second scaling symmetry (\ref{scaling2}) to fix $a_1=1$ and thus $\Gamma=\sigma$. Since $\Gamma>0$ we see that $\sigma>0$. Now, observe that since
$\gamma_{11} \geq 0$ (with equality only possible at isolated points), we must have $\sigma_1^2 \geq  c_2C^{-2}$. In
fact, it is easy to show that the case\footnote{In this case, one
can solve for $c_1=\Lambda c_2C^{-2}$ from which it follows that
$Q=(C^2\sigma^2-c_2)(-\Lambda \sigma C^{-2}-1)$. Therefore if
$\Lambda=0$ there is no root $\sigma_2>\sigma_1$. If $\Lambda<0$
then $\sigma_2=-C^{2}\Lambda^{-1}$, however $Q(\sigma)<0$ for
$\sigma_1<\sigma<\sigma_2$.} $\sigma_1^2=c_2C^{-2}$ (so $c_2>0$) is
incompatible with $\dot{Q}(\sigma_1)>0$ and $\sigma_1<\sigma_2$.
Therefore we must have $\sigma_1^2>c_2C^{-2}$, which implies we have
$\gamma_{11}>0$ everywhere, and therefore the 2-metric $\gamma_{ij}$
degenerates only at the zeroes of $Q(\sigma)$. From the form of the
metric on the horizon it follows that the Killing vectors \be m_i =
d_i\left( \frac{\partial}{\partial x^2}- \omega(\sigma_i)
\frac{\partial}{\partial x^1} \right) \ee for constants $d_i$ and
$i=1,2$ vanish at the degeneration points $\sigma=\sigma_i$.
Further, since $\omega(\sigma_1) \neq \omega(\sigma_2)$ it follows
that $m_1 \neq m_2$. Regularity of the metric on the horizon
requires the orbits of $m_i$ to close in such a way there are no
conical singularities at the points where they vanish. We choose the
constants $d_i$ such that in terms of adapted coordinates defined by
$m_i =\partial / \partial \phi_i$, the periodicity of the orbits is
given by $\phi_i \sim \phi_i+2\pi$. The coordinate transformation
between $(x^1,x^2)$ and $(\phi_1,\phi_2)$ is given by: \be
\label{regcoords} x^1=-[\omega(\sigma_1)d_1
\phi_1+\omega(\sigma_2)d_2\phi_2], \qquad x^2=d_1\phi_1+d_2\phi_2 \;
. \ee To ensure the absence of the conical singularities at
$\sigma=\sigma_1$ and $\sigma=\sigma_2$ one must take \be d_i^2 =
\frac{4\sigma_i(C^2\sigma_i^2-c_2)}{\dot{Q}(\sigma_i)^2} \ee which
therefore determines the $d_i$ up to a sign. The solution is now
globally regular, with $m_1$ vanishing at $\sigma=\sigma_1$ and
$m_2$ vanishing at $\sigma=\sigma_2$. Hence the horizon
$\mathcal{H}$ has $S^3$ topology (or that of a Lens space).

Now we will show that this near-horizon geometry is in fact
isometric to the near-horizon limit of known black holes. In the
$\Lambda=0$ case we will show that it is isometric to the
near-horizon limits of two different known extremal black holes: the
Myers-Perry ($J_1 \neq J_2$) and the slowly rotating KK black hole ($0<G_4J<PQ$). In the
$\Lambda<0$ case we will show it is isometric to the near-horizon
limit of the known extremal rotating AdS$_{5}$ black hole~\cite{HHT} ($J_1 \neq J_2$). We provide the near-horizon
limits of all these black holes in the Appendix.

\paragraph{$\Lambda=0$ case} In this case some of the above formulae simplify. In
particular, using $Q(\sigma_i)=0$ one gets
$C^2\sigma_i^2-c_2=c_1\sigma_i$. Therefore, since above we argued
that $C^2\sigma_i^2-c_2>0$, it follows that $c_1>0$. Then we see
that (\ref{ineq1}) implies $c_2\leq 0$. Further, the fact that $Q$
must have two positive roots requires $c_2<0$ and $c_1^2+4C^2c_2>0$.
Using these results one gets \be d_i^2= \frac{4c_1
\sigma_i^2}{c_1^2+4C^2c_2}, \qquad
\omega(\sigma_i)=\frac{\sqrt{-c_2c_1}}{c_1 C\sigma_i} \; .\ee

In fact, from the results of~\cite{KLR} it is straightforward to
show that this near-horizon geometry is isometric to the
near-horizon limit of the five-dimensional extremal Myers-Perry
solution. To see this, first using the scaling freedom (\ref{scaling1}) to set
$C^2=c_1$ (this can be done as $C^2$ and $c_1$ transform
differently) and hence $c_1+4c_2>0$. Next define two positive
constants $a>b>0$ by: \be a \equiv \frac{1}{\sqrt{c_1}}+
\frac{\sqrt{c_1+4c_2}}{c_1}, \qquad b \equiv \frac{1}{\sqrt{c_1}}-
\frac{ \sqrt{c_1+4c_2}}{c_1} \ee from which it follows that
\begin{equation} C^2 = c_1 = \frac{4}{(a+b)^2}, \qquad c_2 =
-\frac{4ab}{(a+b)^4}, \qquad \sigma_1=\frac{b}{a+b}, \qquad
\sigma_2=\frac{a}{a+b} \; .
\end{equation}
The coordinate change defined by
\begin{equation}
\cos^2\theta =\frac{\sigma-\sigma_1}{\sigma_2-\sigma_1}, \qquad x^1
= \frac{\sqrt{ab}(a+b)^2}{2(a-b)}(\psi - \phi), \qquad x^2 =
\frac{(a+b)}{(a-b)}(b\psi - a\phi) \; ,
\end{equation}
where $0\leq \theta \leq \pi/2$ and $\psi=\phi_1$ and $\phi=\phi_2$,
shows that our near-horizon geometry is identical to that of
extremal Myers-Perry as given in the Appendix in
$(\theta,\psi,\phi)$ coordinates and $(a,b)$ parameters (which is
also the same form as in~\cite{KLR}).

Now we will show how our near-horizon geometry is also isometric to
the near-horizon geometry of the slowly rotating extremal KK black
hole. Define the following positive parameters: \be p \equiv
\frac{1}{C^2} \sqrt{c_1\left(1-\frac{c_2}{C^2} \right)}, \qquad q^2
\equiv \frac{c_1}{c_2^2} \left(1-\frac{c_2}{C^2} \right) , \qquad
\eta^2 \equiv 1+ \frac{4C^2c_2}{c_1^2}\ee so $\eta<1$. It follows
that \be C^2 =\frac{2(p+q)}{(pq)^{3/2}(1-\eta^2)^{1/2}}, \qquad
c_1=\frac{2C^2}{\sqrt{1-\eta^2}} \sqrt{\frac{p}{q}},  \qquad c_2=
-\frac{C^2p}{q} \ee and \be \sigma_1= \sqrt{\frac{p}{q(1-\eta^2)}}
(1-\eta), \qquad \sigma_2= \sqrt{\frac{p}{q(1-\eta^2)}} (1+\eta) \;
.\ee Writing the near-horizon geometry in coordinates
$(\theta,y,\phi)$ defined by: \be \cos\theta =
\frac{2\sigma-\sigma_1-\sigma_2}{\sigma_2-\sigma_1}, \qquad
x^1=-\frac{\sqrt{1-\eta^2}}{C^2 \eta} \phi, \qquad x^2= \frac{2}{C^2
q} \sqrt{ \frac{(p+q)}{p(1-\eta^2)}} \left( \frac{\phi}{\eta} +
\sqrt{\frac{p+q}{p^3}}\; y \right), \ee where $0 \leq \theta \leq
\pi$, shows that it is identical to the near-horizon limit of the
slowly rotating extremal KK black hole given in the Appendix in
$(\theta,y,\phi)$ coordinates and $(p,q,\eta)$ parameters.

\paragraph{$\Lambda<0$ case} Set $\Lambda=-4g^2$. It is
convenient to work with the roots $\sigma_1,\sigma_2,\sigma_3$ of
$Q$ as parameters as well as the original parameters $C^2,c_1,c_2$.
These are related by: \be C^2=4g^2(\sigma_1+\sigma_2+\sigma_3),
\qquad c_1=4g^2(\sigma_1\sigma_2+\sigma_1\sigma_3+\sigma_2\sigma_3),
\qquad c_2=-4g^2 \sigma_1\sigma_2\sigma_3 \ee so
$Q=4g^2(\sigma-\sigma_1)(\sigma-\sigma_2)(\sigma-\sigma_3)$ where
$\sigma_3>\sigma_2$. Define the quantity $W=\frac{\sigma_1\sigma_2+\sigma_1\sigma_3+\sigma_2\sigma_3}{\sigma_1\sigma_2}$ which is invariant under the scaling freedom (\ref{scaling1}). Use the scaling freedom (\ref{scaling1}) to set $\frac{\sigma_3}{\sigma_1\sigma_2}=W$; this can be done as the LHS transforms homogeneously and the RHS is invariant. This implies $\sigma_1+\sigma_2<1$ and \be \sigma_3= \frac{\sigma_1
\sigma_2}{1-\sigma_1-\sigma_2} \;  . \ee
Now define the positive constants
$a,b,r_+$ by: \be \frac{1}{1+g^2r_+^2}= \sigma_1+\sigma_2, \qquad
\frac{r_+^2}{r_+^2+a^2}=\sigma_1, \qquad
\frac{r_+^2}{r_+^2+b^2}=\sigma_2 \ee so $a>b$ (as
$\sigma_1<\sigma_2$). This implies that \be
\sigma_3= \frac{r_+^2(1+g^2r_+^2)}{g^2(r_+^2+a^2)(r_+^2+b^2)} \ee
and \be
C^2=\frac{4r_+^2(1+a^2g^2+b^2g^2+3g^2r_+^2)}{(r_+^2+a^2)(r_+^2+b^2)}
\; . \ee Now define a new variable $\theta$ by \be \cos^2\theta =
\frac{\sigma-\sigma_1}{\sigma_2-\sigma_1} \ee so $0\leq \theta \leq
\pi/2$ uniquely parameterizes the interval $\sigma_1 \leq \sigma
\leq \sigma_2$. This implies \be \Gamma=\sigma= \frac{r_+^2
\rho_+^2}{(r_+^2+a^2)(r_+^2+b^2)}, \qquad Q= \frac{4
r_+^6(a^2-b^2)^2 \sin^2\theta \cos^2\theta
\Delta_{\theta}}{(r_+^2+a^2)^3(r_+^2+b^2)^3} \ee where we have
defined \be \rho_+^2= r_+^2+a^2\cos^2\theta+b^2 \sin^2\theta, \qquad
\Delta_{\theta}=1-a^2g^2 \cos^2\theta -b^2 g^2 \sin^2\theta \; . \ee
It follows that \be \frac{\Gamma d\sigma^2}{Q}= \frac{\rho_+^2
d\theta^2}{\Delta_{\theta}} \; \ee which proves that the
$\sigma\sigma$ component of our horizon metric coincides with the
$\theta\theta$ component of the known extremal rotating AdS$_5$ black hole of~\cite{HHT} (see Appendix). It remains to
check the $x^ix^j$ components of the horizon metric. To do this we
need the constants $d_i,\omega(\sigma_i)$ appearing in the
coordinate transformation (\ref{regcoords}) which work out to be
\bea
d_1 &=& -\frac{r_+^2+b^2}{\Xi_b (a^2-b^2)} \sqrt{ (1+b^2g^2+2g^2r_+^2)(2r_+^2+a^2+b^2)} \\ d_2 &=& \frac{r_+^2+a^2}{\Xi_a (a^2-b^2)} \sqrt{ (1+a^2g^2+2g^2r_+^2)(2r_+^2+a^2+b^2)} \\
\omega(\sigma_1) &=& \sqrt{\frac{(r_+^2+b^2)(r_+^2+a^2)^3(1+a^2g^2+2g^2r_+^2)(1+g^2r_+^2)}{4r_+^4(1+b^2g^2+2g^2r_+^2)(2r_+^2+a^2+b^2)(1+a^2g^2+b^2g^2+3g^2r_+^2)^2}} \\ \omega(\sigma_2) &=& \sqrt{\frac{(r_+^2+a^2)(r_+^2+b^2)^3(1+b^2g^2+2g^2r_+^2)(1+g^2r_+^2)}{4r_+^4(1+a^2g^2+2g^2r_+^2)(2r_+^2+a^2+b^2)(1+a^2g^2+b^2g^2+3g^2r_+^2)^2}}
\eea
where we have defined $\Xi_a=1-g^2a^2$, $\Xi_b=1-g^2b^2$ and without loss of generality we have chosen a particular sign for each of the $d_i$ (note $d_1<0$ and $d_2>0$). Using the transformation (\ref{regcoords}) one can now compute the ${\phi_i \phi_j}$ components of the horizon metric. We have checked that $\gamma_{\phi_i\phi_j}$ is identical to the $a,b= \psi,\phi$ components of the horizon metric of the rotating AdS$_5$ black hole solutions of~\cite{HHT} (see Appendix) upon identifying $\phi_1=\psi$ and $\phi_2=\phi$. Therefore we have verified that the horizon metric of our solutions coincides exactly with that of the known extremal rotating AdS$_5$ black hole. Finally, let us turn to the remaining near-horizon data, the vector $k^i\partial_i=\partial/\partial x^1$. Using the coordinate change (\ref{regcoords})
\bea
\frac{\partial}{\partial x^1} &=& \frac{1}{d_1[\omega(\sigma_2)-\omega(\sigma_1)]} \frac{\partial}{\partial \phi_1} +\frac{1}{d_1[\omega(\sigma_1)-\omega(\sigma_2)]} \frac{\partial}{\partial \phi_1} \\
&=& \frac{2br_+}{\Xi_b(r_+^2+b^2)^2} \frac{\partial}{\partial \phi_1} +\frac{2ar_+}{\Xi_a(r_+^2+a^2)^2} \frac{\partial}{\partial \phi_1}
\eea
where the first equality follows from the coordinate change (\ref{regcoords}) and the second upon using our expressions for $d_i,\omega(\sigma_i)$. Therefore the $k^i$ agree with those of the extremal rotating AdS$_5$ black hole upon the same identification $\phi_1=\psi$ and $\phi_2=\phi$.
Therefore, to summarise, we have proved that $\gamma_{ab},k^i,C^2,\Gamma$ all coincide with those of the most general known extremal rotating AdS$_5$ black hole~\cite{HHT} (as given in the Appendix) thus proving equivalence of the near-horizon geometries.

\subsection{All Ricci flat solutions with compact horizons}
In the $\Lambda=0$ can we can actually determine all possible
near-horizon geometries with compact horizons as we will now show.
Equation (\ref{Gammaode}) integrates to \be Q^3 \frac{d^3
\Gamma}{d\sigma^3} =\alpha \Gamma \ee where $\alpha$ is a constant.
In the Appendix we prove that the LHS is a globally defined function
which vanishes at the zeros of $Q$. Therefore evaluating at one of
the zeros of $Q$ implies that $\alpha=0$. It follows that \be
\frac{d^3 \Gamma}{d\sigma^3}=0 \ee and therefore \be \Gamma=
a_2\sigma^2+a_1\sigma+a_0 \ee where $a_i$ are integration constants.
Also,  equation (\ref{ddQ}) determines $Q$: \be Q=-C^2 \sigma^2 +c_1
\sigma +c_2 \ee where $c_1,c_2$ are constants. The analysis now
splits into two cases: either $a_2=0$ or $a_2 \neq 0$. We have
already analysed the former case in the previous section where it was shown that the resulting near-horizon geometry is identical to the near-horizon limit of extremal Myers-Perry, or equivalently the near-horizon limit of the slowly rotating extremal KK black hole.

We now analyse the $a_2 \neq 0$ case. Since $\sigma$ is only defined up to an
additive constant, we can always shift $\sigma$ to set  $a_1=0$ and
thus without loss of generality we take \be \Gamma=a_2 \sigma^2+a_0
\; . \ee Substituting into the equation for $k^ik_i$ (\ref{kiki})
gives \be \gamma_{11}= \frac{2P(\sigma)}{\Gamma} \ee where we have
defined \be P(\sigma) \equiv  \alpha \sigma^2+ \beta\sigma +\gamma
\ee and \be \alpha=-C^2a_0a_2-c_2a_2^2, \qquad \beta =2a_0a_2c_1,
\qquad \gamma= C^2a_0^2+a_2a_0c_2 \ee which satisfy $\gamma
a_2+\alpha a_0=0$ and the discriminant of the quadratic $P$ is \be
\label{D} D \equiv
\beta^2-4\alpha\gamma=4a_0a_2[c_1^2a_0a_2+(C^2a_0+a_2c_2)^2] \; .
\ee Now, plugging into (\ref{omegadot1}) gives \be \label{pos}
\dot{\omega}^2 = \frac{(a_0C^2-a_2c_2) [ c_1^2a_0a_2
+(C^2a_0+a_2c_2)^2] \Gamma^2}{2P(\sigma)^4} \; . \ee
Notice, that this implies the constants satisfy \be \label{constineq}
(a_0C^2-a_2c_2)[ c_1^2a_0a_2 +(C^2a_0+a_2c_2)^2] \geq 0 \; . \ee
The analysis thus splits into a number of subcases. In the Appendix we
show that $[ c_1^2a_0a_2 +(C^2a_1+a_2c_2)^2]=0$ does not lead to a
compact horizon and therefore we exclude this. It follows that there
are two possibilities (i) $a_0C^2-a_2c_2=0$ or (ii) $a_0C^2-a_2c_2
\neq 0$.

\subsubsection{Inhomogeneous $S^1 \times S^2$ horizon} We now consider case (i) and eliminate $a_0$ using $a_0=a_2c_2C^{-2}$. Observe that (\ref{pos}) implies $\omega$ is a constant. Also note that in this case the  quadratic $P(\sigma) \propto Q(\sigma)$; in particular
\be \gamma_{11}= \frac{4c_2a_2^2 Q(\sigma)}{C^2 \Gamma} \; . \ee The
horizon metric reads \be \gamma_{ab}dx^a dx^b = \frac{a_2(
c_2C^{-2}+\sigma^2)}{Q(\sigma)} d\sigma^2 + \frac{4c_2 a_2}{C^2
(c_2C^{-2} +\sigma^2)} Q(\sigma) (dx^1+\omega dx^2)^2 +
\frac{C^2}{4c_2a_2^2}(dx^2)^2 \ee with $\Gamma= a_2( c_2C^{-2}
+\sigma^2)$. This metric is non-degenerate everywhere except at the
end points $\sigma=\sigma_1$ and $\sigma=\sigma_2$ where $Q=0$. At
these points $\partial / \partial x^1$ vanishes and the metric as
conical singularities in general. The simultaneous removal of these
conical singularities leads to a regular metric on $S^1 \times S^2$.
The condition for this is easily shown to be \be
\frac{\dot{Q}(\sigma_1)}{\Gamma(\sigma_1)}=-\frac{\dot{Q}(\sigma_2)}{\Gamma(\sigma_2)}
\ee which noting $\dot{Q}(\sigma_i)=\mp C^2(\sigma_1-\sigma_2)$
implies $\Gamma(\sigma_1)=\Gamma(\sigma_2)$. It follows that
$\sigma_2=-\sigma_1$ and hence $c_1=0$ and $c_2>0$. Since $\Gamma>0$, now it follows that $a_2>0$.  Now, rescaling
$\sigma \to \sqrt{c_2}C^{-1} \sigma$ and $x^2 \to C c_2^{-1/2} x^2$
and defining a new coordinate and parameter by
\be \phi =C^2 x^1, \qquad a\equiv \frac{\sqrt{a_2c_2}}{C^2} \ee one finds \bea
\gamma_{ab}dx^adx^b &=& \frac{a^2(1+\sigma^2)}{1-\sigma^2} d\sigma^2 +
\frac{4a^2(1-\sigma^2)}{(1+\sigma^2)} (d\phi+ \Omega dx^2)^2+ \frac{1}{4C^4a^4}(dx^2)^2 \eea where $\Gamma=C^2a^2(1+\sigma^2)$ and we have defined a new constant $\Omega \equiv \omega
C^3c_2^{-1/2}$. The Killing vector
$k=\partial/\partial x^1=C^2 \partial/\partial \phi$ vanishes at $\sigma =\pm 1$; absence of conical singularities at these points implies $\phi\sim \phi +2\pi$ and therefore $\partial /\partial \phi$ generates a rotational symmetry. Finally, we use the shift freedom $\phi \to \phi +\const x^2$ in order to ensure $\partial /\partial x^2$ corresponds to the other rotational symmetry generator, so $x^2 \sim x^2 +L$. We have thus derived a near-horizon geometry whose horizon topology is $S^1 \times S^2$. It is parameterized by $(a,C,\Omega,L)$, although there is a scaling symmetry
\be
C^2 \to KC^2, \qquad \Omega \to K^{-1}\Omega, \qquad L\to KL, \qquad x^2 \to Kx^2
\ee
which allows one to fix a combination of $(C,\Omega,L)$ (note $a$ is invariant) and hence it is a three parameter family.

In fact, in a particular region of the parameter space, the above
near-horizon geometry is isometric to that of the extremal boosted
Kerr string. This region is given by $C^2|\Omega|<1/(4a^3)$ (which
is invariant under the scaling symmetry above). In this region
define a boost parameter $\beta$ (invariant under the scaling
symmetry) by $\tanh \beta\equiv  4a^3 C^2\Omega$. Then use the
scaling freedom to set $C^2=1/(2a^2 \cosh \beta)$ and thus one can
solve for $\Omega=(\sinh \beta)/(2a)$. Changing coordinates to
$\sigma=\cos\theta$, with $0\leq \theta \leq \pi$, we see that this
near-horizon geometry is identical to that of the extremal boosted
Kerr-string as given in~\cite{KLR}. Note that the special case
$\sinh^2\beta=1$ corresponds to the near-horizon geometry of the
asymptotically flat extremal vacuum black ring~\cite{PS} as first
observed in~\cite{KLR}.  It is curious that the boosted Kerr string
``misses'' the region of parameter space given by $C^2|\Omega|\geq
1/(4a^3)$.

\subsubsection{Inhomogeneous $S^3$ horizon}
 We now analyse case (ii), i.e. $a_0 \neq a_2c_2 C^{-2}$. It proves
convenient to split the analysis into two cases depending on whether
$\alpha=0$ or not. First consider $\alpha \neq 0$. Integrating (\ref{pos}) gives
\be \label{omegaflat} \omega= \pm \left[ -\frac{ \kappa a_2\sigma
}{\alpha P(\sigma)} +c_3 \right]\ee where for convenience we have defined a constant $\kappa > 0$ by
\be \kappa \equiv \sqrt{ \frac{(a_0C^2 - a_2c_2)[
c_1^2a_0a_2 +(C^2a_0+a_2c_2)^2]}{2}} \; \ee and $c_3$ is an
integration constant. The remaining equations are satisfied without
further constraint.

We will use the shift freedom $x^1 \to x^1 +\const x^2$ to set $c_3=0$ and $x^2 \to \pm x^2$ to pick a sign for $\omega$. The horizon metric is
\be \gamma_{ab}dx^a dx^b = \frac{\Gamma d\sigma^2}{Q(\sigma)}+
\frac{2P(\sigma)}{\Gamma}\left[ dx^1 - \frac{\kappa a_2\sigma}{\alpha P(\sigma)}
dx^2 \right]^2 + \frac{Q(\sigma)}{2P(\sigma)} (dx^2)^2 \; .\ee The following
identity is easily verified \be \label{Pident} P(\sigma) \equiv
(C^2a_0-a_2c_2)\Gamma(\sigma)+2a_0a_2Q(\sigma) \; , \ee which
implies $P(\sigma_i)=(C^2a_0-a_2c_2)\Gamma(\sigma_i)$. For a
positive definite metric we must have $P(\sigma_i) \geq 0$, which
implies $a_0 > a_2c_2C^{-2}$ and thus $P(\sigma_i)>0$. Observe that from (\ref{constineq})
it follows that $[ c_1^2a_0a_2
+(C^2a_1+a_2c_2)^2] > 0$. There are now two cases to consider:
either the discriminant $D \geq 0$ or $D<0$. Using (\ref{D}) we see
that $D \geq 0$ is then equivalent to $a_0a_2 \geq 0$ and $D<0$ is
equivalent to $a_0a_2<0$. Therefore, in the case $D \geq 0$,
equation (\ref{Pident}) implies $P(\sigma)>0$ for $\sigma_1 \leq
\sigma \leq \sigma_2$. On the other hand, if $D<0$, in which case
$P$ has no real roots, then it must be the case that $P(\sigma)>0$
for all $\sigma$ (so $\alpha>0$).  Therefore we see that in both
cases $P>0$ for $\sigma_1 \leq \sigma \leq \sigma_2$ and therefore
the metric on the horizon is non-degenerate everywhere except at the
endpoints $\sigma_1,\sigma_2$ where $Q$ vanishes. The Killing
vectors \be \label{u1coords} m_i =d_i \left( \frac{\partial}{\partial x^2}
-\omega(\sigma_i) \frac{\partial}{\partial x^1} \right) \ee for
constant $d_i$ vanish at the endpoints $\sigma=\sigma_i$, where the
metric has conical singularities in general. Using $Q(\sigma_i)=0$
it can be shown that $\omega(\sigma_1) \neq \omega(\sigma_2)$ and
therefore $m_1 \neq m_2$. Thus, removing the conical singularities
(which corresponds to a particular choice of $d_i$) gives a metric
which $S^3$ topology. The values of $d_i$ work out to be: \be d_i^2
= \frac{8 P(\sigma_i) \Gamma(\sigma_i)}{\dot{Q}(\sigma_i)^2}=
\frac{8(C^2 a_0-a_2c_2)
\Gamma(\sigma_i)^2}{C^4(\sigma_1-\sigma_2)^2} \; . \ee

Now let us consider the $\alpha=0$ case. In the Appendix we show that this arises as a limit of the $\alpha \neq 0$ case. In fact in the appendix we give expressions valid for $\beta \neq 0$ which maybe be viewed as complementary to the $\alpha \neq 0$ case, since one cannot have both $\alpha=\beta=0$ (as then $P \equiv 0$).

The near-horizon metric has the following scaling symmetries (corresponding to $\mathcal{S}_2^{-1}\mathcal{S}_1$ and $\mathcal{S}_2$):
\bea \label{scale1} &&C^2 \to K C^2, \quad a_0
\to Ka_0 \quad a_2 \to K a_2, \quad c_1 \to Kc_2 \quad c_2 \to Kc_2
\nonumber
\\ && x^1\to K^{-1}x^1, \quad x^2 \to K x^2, \quad v \to K^{-1} v
\eea
where $K>0$ and
\be \label{scale2} a_2 \to L^{-2}a_2, \qquad c_1 \to L c_1, \qquad c_2 \to L^2 c_2, \qquad \sigma \to L \sigma, \qquad x^2 \to L^{-1} x^2
\ee
where $L$ is a constant (of either sign).
These may be used to fix two (or two combinations) of the parameters $(a_0,a_2,c_1,c_2,C^2)$. Therefore this is a 3 parameter family of solutions.

We will now show that in a particular region of parameter space the $a_2>0$ solution is isometric to the near-horizon geometry of the fast rotating extremal KK black hole (i.e. $G_4J>PQ$). Observe that $X \equiv
\frac{(c_1^2+4C^2c_2)}{4C^4}$ is invariant under the first symmetry (\ref{scale1})
and scales as $X \to L^2X$ under the second symmetry (\ref{scale2}). Therefore, use the second symmetry
to set $X=1$. Note that since the condition $X=1$ is invariant under the first
symmetry, we are still free to use (\ref{scale1}). Define a positive constants
$p,Z$ by\be
\label{pZdef} p^2 \equiv \frac{c_1^2a_0a_2+(C^2a_0+a_2c_2)^2 }{C^6a_2}, \qquad  Z\equiv \frac{4a_2}{C^2}+\frac{2}{C^4}(C^2a_0-a_2c_2) \; .\ee  Note that $p$ and $Z$ are invariant under (\ref{scale1}). There are now two
possibilities: either $Z/p^2<1$ or $Z/p^2 \geq 1$. The former region
of parameter space gives the fast KK black hole as we now show. Use
the first symmetry (\ref{scale1}) to set \be \label{sym1} C^2a_0-a_2c_2=
\frac{C^2}{2a_2}\left(1-\frac{Z}{p^2} \right) \ee which is possible as the LHS
transforms homogeneously (i.e. as $K^2$), but the RHS is invariant, and also for our solution $C^2a_0-a_2c_2>0$.
We also define positive constants $a,q$ by
\be
a^2 \equiv \frac{a_2}{C^2}, \qquad  q \equiv \frac{1}{pC^2a_2} \ee which can
be inverted to give
\be C^2=\frac{1}{a\sqrt{pq}}, \qquad  a_2 = \frac{a}{\sqrt{pq}}. \ee Now,
using (\ref{sym1}) it follows that \be \label{imp} C^2a_0-a_2c_2 =
\frac{p^2-4a^2}{2a^2p(p+q)} \ee and thus $p^2-4a^2>0$. Note that
using $X=1$, (\ref{pZdef}) can be written as $p^2\equiv
\frac{4a_0}{C^2}+\frac{1}{C^6a_2}(C^2a_0-a_2c_2)^2$; this, together
with (\ref{imp}) can then be used to solve for $a_0$ to give \be
a_0=\frac{1}{a\sqrt{pq}}\left( \frac{p^2}{4}-
\frac{q^2(p^2-4a^2)^2}{16a^2(p+q)^2} \right) \; .\ee  Then
(\ref{imp}) can be used to solve for $c_2$ giving: \be c_2=
\frac{1}{a\sqrt{pq}}-\frac{p^2(p^2-4a^2)(q^2-4a^2)}{16a^5\sqrt{pq}
(p+q)^2} \; .\ee  Finally use $X=1$ to solve for $c_1^2$:
\be c_1^2=\frac{p(p^2-4a^2)(q^2-4a^2)}{4a^6q(p+q)^2} \ee which
implies $q^2 \geq 4a^2$. Thus $c_1$ is determined up to a sign. To
fix the sign recall that when we used the second
symmetry to set $X=1$ we did not specify the sign of $L$; therefore
we can use this sign freedom to ensure $c_1>0$. Using the scaling symmetries, we have therefore shown how to go between the two sets of parameters $(C^2,a_0,a_2,c_1,c_2)$ and $p,q,a$ in the region defined by $a_2>0$ and $Z<p^2$.

Now, define a new coordinate by
\begin{equation}
\cos\theta = \frac{2\sigma - \sigma_1-\sigma_2}{\sigma_2-\sigma_1}=\sigma-\frac{c_1}{2C^2}.
\end{equation}
so $0\leq \theta \leq \pi$ uniquely parameterizes the interval $\sigma_1\leq \sigma \leq \sigma_2$. This implies
\be
Q(\sigma)=C^2 \sin^2\theta, \qquad \Gamma=C^2H_p
\ee
where $H_p$ is defined in (\ref{Hfast}) from which it follows
\be
\frac{\Gamma d\sigma^2}{Q(\sigma)}= H_p d\theta^2 \; .
\ee
This proves that the $\sigma\sigma$ component of our near-horizon geometry written in the $\theta$ coordinate introduced agrees with the $\theta\theta$ component of the near-horizon limit of the  fast rotating extremal KK black hole as given in the Appendix. In order to verify the rest of the horizon metric we need to evaluate the constants $d_i$ and $e_i \equiv -d_i \omega(\sigma_i)$ appearing in the coordinate transformation defined by (\ref{u1coords}) and $m_i =\partial /\partial \phi_i$. One finds\footnote{The expression for $e_i$ is valid for $\alpha \neq 0$. For $\alpha=0$ one must use an expression for $\omega$ valid when $\alpha=0$. This is given in the Appendix and simply amounts to a shift in $\omega$ which can be generated by shifting $x^1$. It thus suffices to check $\alpha \neq 0$ as then the $\alpha=0$ case follows by the appropriate shift in $x^1$.}
\be
d_i=\epsilon_i \sqrt{\frac{q(p^2-4a^2)}{p+q}} \Gamma(\sigma_i), \qquad e_i=-\frac{\epsilon_i a \sigma_i}{\alpha q}
\ee
where
\begin{eqnarray}
\Gamma(\sigma_1) &=& \frac{p}{2a\sqrt{pq}(p+q)}\left[(pq + 4a^2) -  \sqrt{(p^2-4a^2)(q^2-4a^2)}\right] \nonumber \\ \Gamma(\sigma_2) &=& \frac{p}{2a\sqrt{pq}(p+q)}\left[(pq + 4a^2) + \sqrt{(p^2-4a^2)(q^2-4a^2)}\right]
\end{eqnarray}
and we will chose the signs by $\epsilon_1=+ 1$ and $\epsilon_2=-1$. In fact the KK black hole is usually written in ``Euler'' type coordinates which are related to the $\phi_i$ by $\phi=\phi_1+\phi_2$ and $y=2P(\phi_2-\phi_1)$ where $P\equiv \sqrt{\frac{p(p^2-4a^2)}{4(p+q)}}$. It is thus convenient to change coordinates directly from $x^i$ to $(y,\phi)$ which is performed by
\be
x^1=\frac{1}{2}(e_1+e_2)\phi+ \frac{1}{4P}(e_2-e_1)y, \qquad x^2=\frac{1}{2}({d}_1+{d}_2)\phi+({d}_2-{d}_1)\frac{y}{4P} \; .
\ee
We have checked that for our near-horizon geometry $\gamma_{ij}$ written in $(y,\phi)$ coordinates coincides with the $(y,\phi)$ components of the horizon metric of the fast rotating KK black hole given in the Appendix. Furthermore, using the coordinate transformation above, one can calculate $k^y$ and $k^{\phi}$ (recall $k^1=1,k^2=0$) which also coincide with those of the fast rotating KK black hole given in the Appendix. This ends the proof of the equivalence of our $a_2>0$ near-horizon geometry in the region of parameter space defined by $Z<p^2$ (where $Z$ and $p$ are defined as in (\ref{pZdef})) and the near-horizon limit of the fast rotating KK black hole.

\subsection{Near-horizon geometry of a ``small'' extremal $AdS_5$ black ring?}
\label{smallring}
We have not been able to solve the near-horizon equations in general in $D=5$ with $\Lambda<0$. Earlier we showed that to do this one needs to solve the two coupled ODEs (\ref{ddQ}) and (\ref{Gammaode}) for $Q$ and $\Gamma$. We first note that when $\Lambda \neq 0$ it is possible to eliminate $\Gamma$ from (\ref{Gammaode}) using (\ref{ddQ}), resulting in a 6th order ODE for $Q$:
\begin{equation}\label{6ODE}
\frac{d}{d\sigma} \left( \frac{Q^3}{\ddot{Q}+2C^2} \frac{d^5 Q}{d\sigma^5} \right) + \frac{5}{3} Q^2 \frac{d^4Q}{d\sigma^4}=0.
\end{equation} Given a solution to this one can then deduce $\Gamma$ from (\ref{ddQ}). Finding all solutions to (\ref{6ODE}) would lead to the classification of all allowed near-horizon geometries of extremal vacuum black holes with $R \times U(1)^2$ symmetry in $AdS_{5}$. Curiously all explicit dependence in $\Lambda$ has cancelled from this 6th order ODE (although we emphasise it is only valid when $\Lambda \neq 0$) -- it is thus more convenient to work with the coupled pair of ODEs (\ref{ddQ}) and (\ref{Gammaode}). We will now present some results which follow from these equations. \\

\noindent {\it Lemma}: The most general polynomial solution to (\ref{ddQ}) and (\ref{Gammaode}) is given by $\Gamma=a_0+a_1 \sigma$ and $Q=-\Lambda a_1 \sigma^3 -
(C^2+3\Lambda a_0)\sigma^2 +c_1 \sigma +c_2 $. \\

\noindent {\it Proof:} First observe that (\ref{ddQ}) implies that $Q$ is a polynomial iff $\Gamma$ is a polynomial. Suppose $\Gamma$ is a polynomial or order $n = 2$. The ODE (\ref{Gammaode}) then implies $Q^2=0$ and then (\ref{ddQ}) implies $\Gamma=0$, a contradiction. Now suppose $\Gamma$ is a polynomial or order $n \geq 3$. For $\sigma \to \infty$ we have $\Gamma \sim a_n \sigma^n$ for some non-zero constant $a_n$. The ODE (\ref{ddQ}) then implies $Q \sim -6\Lambda a_N \sigma^{n+2}/[(n+1)(n+2)]$. Then, examining the $\sigma \to \infty$ limit of the ODE (\ref{Gammaode}) implies $n=4/7$ which is a contradiction. This leaves $\Gamma=a_1\sigma+a_0$ which is indeed a solution with the $Q$ given above.\\

As we showed in an earlier section $\Gamma=a_1\sigma+a_0$ gives the near-horizon geometry of the known extremal rotating AdS$_5$ black hole~\cite{HHT}, which has spherical horizon topology. An interesting question is whether there exists a near-horizon geometry with $S^1 \times S^2$ topology thus providing a candidate extremal AdS black ring near-horizon geometry. Recall that the near-horizon limit of the asymptotically flat black ring has $\Gamma=a_0+a_2\sigma^2$. But the above Lemma tells us that this cannot be the case when one has a cosmological constant. This is perhaps surprising as the near-horizon limits of the topologically spherical Myers-Perry black hole and its generalization to include a negative cosmological constant both have $\Gamma$ of the same form (a linear polynomial).

If an extremal vacuum AdS black ring does exist, one might expect it to be continuously connected to the asymptotically flat extremal vacuum black ring as one turns off the cosmological constant. It is thus of interest to investigate the existence of ``small'' AdS black rings, in the sense that both the radius of the $S^1$, say $R_1$, and the $S^2$, say $R_2$, are much smaller than the AdS length scale $\ell$ ($\Lambda=-4/\ell^2$). For the asymptotically flat extremal black ring $R_2 \sim a$ (where $a$ is the Kerr parameter in the corresponding boosted Kerr string solution) and $R_1$ is just proportional to the period of $z$ (which does not appear explicitly in the near-horizon geometry, only implicitly through identification of $z$). Therefore we will consider linearising the pair of ODEs about the solution corresponding to the boosted Kerr string near-horizon geometry (which includes that of the extremal black ring) for small $a/\ell$ (or equivalently small $\Lambda C^{-2})$. In our formalism a near-horizon geometry is specified by the data $(C^2,\Gamma,Q,\gamma_{ij})$ (recall we set $k^1=1,k^2=0$) and thus this is the data which we must linearise about.

Expand\footnote{Note that for the class of $SO(2,1)\times
U(1)^2$-invariant near-horizon geometries we have been considering,
$\sigma,\Gamma,Q,C^2$ are invariantly defined quantities up to the
two constant scaling symmetries (\ref{S1}), (\ref{S2}). Therefore
the only gauge freedom in our perturbation analysis are these
constant scalings. These can be fixed by working with a background solution in which the scaling symmetries have been used to fix the parameterisation, as for the boosted Kerr-string. }\be Q(\sigma)= Q_0(\sigma)+ \epsilon Q_1(\sigma)
+O(\epsilon^2), \quad \Gamma(\sigma)= \Gamma_0(\sigma)+ \epsilon
\Gamma_1(\sigma) +O(\epsilon^2), \quad C^2=C_0^2(1+ A_1\epsilon
+O(\epsilon^2)) \ee where $\epsilon \equiv \Lambda C^{-2}_0$ is a
dimensionless expansion parameter, $Q_1,\Gamma_1$ and $A_1$ are
dimensionless functions and constant respectively, and \be
Q_0=C_0^2(1-\sigma^2), \qquad
\Gamma_0=\frac{(1+\sigma^2)}{2c_{\beta}} ,\qquad C_0^2=\frac{1}{2a^2
c_{\beta}} \ee is the zeroth order data corresponding to the Kerr
string (which we denote with a $0$ subscript). Plugging this into
the ODEs (\ref{ddQ}) and (\ref{Gammaode}) gives : \be
\ddot{Q_1}+2C_0^2A_1+6 \Gamma_0 C_0^2=0, \qquad \frac{d}{d\sigma}
\left( \frac{Q_0^3}{\Gamma_0} \frac{d^3 \Gamma_1}{d\sigma^3}
\right)-\frac{10C_0^2Q_0^2}{c_{\beta}} =0 \ee which determines $Q_1$
and $\Gamma_1$. Explicitly \be Q_1= C_0^2\left[-\frac{\sigma^4}{4
c_{\beta}} - \left(A_1+\frac{3}{2c_{\beta}}
\right)\sigma^2+d_1\sigma+d_2 \right] \ee where $d_1,d_2$ are
integration constants and \be \frac{d^3\Gamma_1}{d\sigma^3} =
\frac{10C_0^2\Gamma_0}{c_{\beta}Q_0^3} \int^{\sigma} d\sigma'
Q_0(\sigma')^2 \ee which determines $\Gamma_1$ up to quadratures. We
find \bea
\Gamma_1 &=& -\frac{\sigma^4}{24 c_{\beta}^2} +\frac{\sigma^2}{2}\left( \frac{1}{c_{\beta}^2}+e_2 \right)+\sigma \left( e_3-\frac{5e_1}{8C_0^4c_{\beta}^2} \right)+e_4 \nonumber \\
&&+ \frac{(1+\sigma^2)}{2}\left[ \left( \frac{5e_1}{8 C_0^4c_{\beta}^2}-\frac{1}{3c_{\beta}^2} \right)\log(1+\sigma) +\left( -\frac{5e_1}{8 C_0^4c_{\beta}^2}-\frac{1}{3c_{\beta}^2} \right)\log(1-\sigma) \right]
\eea
where $e_i$ are four constants of integration. Now, using (\ref{kiki}) we may compute  $\gamma_{11}$ to linear order in $\epsilon$, which for later convenience we write as
\bea
\label{gam11F}
\gamma_{11} &=& \frac{Q}{c_{\beta}^2 \Gamma} \left[1+\epsilon F+ O(\epsilon^2) \right], \\
F&=&\frac{\left[ 2\sigma^4+ \sigma^2( 6c_{\beta}(A_1-d_2)-6c_{\beta}^2(2e_4+3e_2) -17) +11+6c_{\beta}(A_1-d_2) +6c_{\beta}^2(e_2+6e_4) \right]}{12c_{\beta}(1-\sigma^2)} \nonumber \\ \nonumber
&&+ \frac{(15e_1-8C_0^4)}{12c_{\beta}C_0^4} \log(1+\sigma)-\frac{(15e_1+8C_0^4)}{12c_{\beta}C_0^4} \log(1-\sigma) \; .
\eea
We now turn to determining $\omega= \gamma_{12}/\gamma_{11}$. Equation (\ref{omegadot1}) determines $\dot{\omega}^2$ in terms of $(\Gamma,Q,\gamma_{11})$ which can now calculated to linear order in $\epsilon$. Recall that we also showed that $\dot{\omega}\gamma_{11}^2 \Gamma =k$, where $k$ is a constant (\ref{intcond}).
Using (\ref{omegadot1}) we compute this quantity to linear order and find
\be
\label{kconst}
\dot{\omega}^2\gamma_{11}^4\Gamma^2= \frac{4\epsilon C_0^6}{3c_{\beta}^4} \left[ 3c_{\beta} (A_1-d_2)+3 c_{\beta}^2(-e_2+2e_4) -1\right] +O(\epsilon^2)
\ee
which is indeed a constant; in fact note that for generic parameter values $k=O(\sqrt{\epsilon})$.  Integrating for $\omega$ gives
\be
\omega= k\left( \frac{\sigma}{1-\sigma^2} + O(\epsilon) \right) +\omega_0
\ee
where the constant $\omega_0$ is the $\epsilon=0$ value of the boosted Kerr string.

Let us now analyse regularity of this perturbative solution. First, observe that the location of the roots of $Q$ change, so write them as $\sigma_{\pm}= \pm 1 + \epsilon \delta \sigma_{\pm} + O(\epsilon^2)$, where we have written $\sigma_+=\sigma_2$ and  $\sigma_-=\sigma_1$ for convenience. Inserting into $Q$ gives:
\be
\delta \sigma_{\pm}=\pm \frac {Q_1(\pm1)}{2C_0^2} =\pm\frac{1}{2}\left( -\frac{7}{4c_{\beta}} +d_2-A_1 \pm d_1 \right)
\ee
and regularity requires $\delta \sigma_+ >0$ and $\delta \sigma_{-}<0$ to ensure that $\log(1\pm \sigma)$ is regular in the relevant interval $[\sigma_-,\sigma_+]$ (note $\epsilon<0$). For consistency of our perturbation series we require that the various metric functions evaluated at the endpoints $\sigma_{\pm}$ coincide with those of the boosted Kerr string as $\epsilon \to 0$. It turns out, $\Gamma(\sigma_{\pm})=\frac{1}{c_{\beta}}+ O(\epsilon \log |\epsilon |)$ due to the logarithm terms. However, the function $F$ (appearing in $\gamma_{11}$) and $\omega$ both contain factors of $1/(1-\sigma^2)$ which at the end points contribute $O(\epsilon^{-1})$ -- as a result for generic parameter values $F(\sigma_{\pm})=O(\epsilon^{-1})$ and $\omega=O(\epsilon^{-1/2})$. Both of these are not acceptable: we must choose parameters such that the factor of $1-\sigma^2$ in the denominator of the first term of $F$ cancels with its numerator, and also impose that the constant $k= O(\epsilon)$ so $\omega=O(1)$. Demanding that the $O(\epsilon)$ term in (\ref{kconst}) vanishes gives
\be
\label{A1}
A_1-d_2=\frac{1}{3c_{\beta}}-c_{\beta}(2e_4-e_2)
\ee
which is equivalent to $k=O(\epsilon)$. Using this to eliminate $A_1-d_2$ in $F$ (\ref{gam11F}) implies, remarkably, that the numerator of the first term has a factor of $1-\sigma^2$ which thus cancels the unwanted factor in the denominator leaving
\be
F=c_{\beta}(e_2+2e_4)+\frac{13-2\sigma^2}{12 c_{\beta}}+ \frac{(15e_1-8C_0^4)}{12c_{\beta}C_0^4} \log(1+\sigma)-\frac{(15e_1+8C_0^4)}{12c_{\beta}C_0^4} \log(1-\sigma) \; .
\ee
Therefore, we have $\epsilon F(\sigma_{\pm})= O(\epsilon \log |\epsilon|)$. We conclude that the near-horizon solution we have is valid to order $O(\epsilon^2)$ for $\sigma_- \leq \sigma \leq \sigma_+$.

We must also ensure the absence of conical singularities in the horizon metric which reads
\be
\gamma_{ab}dx^a dx^b = \frac{\Gamma d\sigma^2}{Q} + \frac{Q}{c_{\beta}^2\Gamma}( 1+\epsilon F +O(\epsilon^2))(dx^1+\omega dx^2)^2 + c_{\beta}^{2}( 1-\epsilon F +O(\epsilon^2))(dx^2)^2 \; .
\ee
Simultaneous removal of conical singularities is equivalent to
\be
\frac{\dot{Q}(\sigma_+)(1+\frac{\epsilon }{2}  F(\sigma_+) + O(\epsilon^2))}{\Gamma(\sigma_+)} =-\frac{\dot{Q}(\sigma_-)(1+\frac{\epsilon }{2} F(\sigma_-) + O(\epsilon^2))}{\Gamma(\sigma_-) } \; .
\ee
It is easily seen that this can be satisfied if $Q,F,\Gamma$ are even\footnote{Consider a metric of the form $\gamma_{ab}dx^a dx^b = \frac{\Gamma d\sigma^2}{Q} + \frac{QP}{\Gamma }(dx^1+\omega dx^2)^2 + P^{-1}(dx^2)^2$ with $\Gamma,Q,P$ functions of $\sigma$ and $\Gamma,P>0$ with $Q$ having two distinct zeros $\pm \sigma_0$ with $Q>0$ in between. The condition for simultaneous removal of the conical singularities at $\sigma=\pm \sigma_0$ (at which points $\partial /\partial x^1$ vanishes) is $-\dot{Q}(\sigma_0)P(\sigma_0)^{1/2}\Gamma(\sigma_0)^{-1}= \dot{Q}(-\sigma_0)P(-\sigma_0)^{1/2}\Gamma(-\sigma_0)^{-1}$. Clearly, this is satisfied if $Q,\Gamma,P$ are even functions of $\sigma$. Then, in the interval $-\sigma_0\leq \sigma \leq \sigma_0$ it is a smooth metric on $S^2 \times S^1$.} functions in $\sigma$. This can be achieved by setting $d_1=e_1=e_3=0$.

With the choices the various functions simplify:
\bea
\Gamma_1 &=& -\frac{\sigma^4}{24 c_{\beta}^2} +\frac{\sigma^2}{2}\left( \frac{1}{c_{\beta}^2}+e_2 \right)+e_4 \nonumber - \frac{(1+\sigma^2)}{6c_{\beta}^2}\log(1-\sigma^2), \\
F &=&c_{\beta}(e_2+2e_4)+\frac{13-2\sigma^2}{12 c_{\beta}}-
\frac{2}{3c_{\beta}}\log(1-\sigma^2) \; . \eea Note that determining
$\omega$ (i.e. the constant $k$) requires a higher order
calculation: one needs the $O(\epsilon^2)$ term in (\ref{kconst})
which we will not pursue here. The perturbation we have constructed
is parameterized by $e_2,e_4,d_2$ (on top of three parameters of
boosted Kerr string) with $A_1$ determined by (\ref{A1}) and
$2e_4-e_2>\frac{25}{12c_{\beta}^2}$ (this is equivalent to the $\pm
\delta \sigma_{\pm}>0$ condition).

For a boost value given by $\sinh^2\beta=1$ the near-horizon
geometry of the Kerr string is isometric to that of the
asymptotically flat extremal black ring which is a 2-parameter
family of solutions (these can be taken to be the two angular momenta $J_i$). One would expect an AdS extremal black ring to
also have 2-parameters. However, the regular perturbations we have
derived depend on more parameters. Presumably these extra parameters
must be fixed somehow (perhaps asymptotic information) for our
perturbative solution to be interpreted as the near-horizon geometry
of a ``small'' AdS black ring.

We can introduce coordinates $(\phi,z)$ where $\phi = d_1 x^{1}$
where $d_1$ is chosen to ensure $\phi$ has period $2\pi$, and
$z=x^2$ runs along a periodic direction (corresponding to that of
the string in the unperturbed case). As explained
in~\cite{KLR,FKLR}, we expect $\partial_\psi$ generating the $S^1$
of the presumptive black ring solution to be given by a linear
combination of $\partial_{\phi}$ and $\partial_z$, while
$\partial_{\phi}$ can be taken to be the generator of the $U(1)$ in
the transverse $S^2$. From our linearized solution above, we can
readily compute $J_\phi$ via a Komar integral~\cite{FKLR}. However,
to determine $J_{z}$ and hence $J_{\psi}$, we require knowledge of
the $O(\epsilon^2)$ term in (\ref{kconst}) which is not available
from our first order calculation.
Physically, one would expect that a black ring in $AdS_{5}$ would have greater angular momenta in the $S^1$ direction, relative to the corresponding asymptotically flat solution, in order to prevent self-collapse.

To summarise we have constructed an approximate solution to the vacuum near-horizon equations with a negative cosmological constant by perturbing about the near-horizon geometry of the boosted Kerr-string. To this level of approximation it describes a regular near-horizon geometry with horizon topology $S^1 \times S^2$. Taking the boost to be that of the asymptotically flat black ring $\sinh^2\beta=1$ provides a candidate for a near-horizon geometry of a ``small'' extremal ring in $AdS_5$.

\section{Discussion}
In this paper we have shown how one may determine all possible
vacuum near-horizon geometries of extremal (but nonsupersymmetric)
black holes under in 4d and 5d under the following assumptions. In 4d we assume
axisymmetry and that the horizon has compact sections of non-toroidal topology. In 5d we assume there are two commuting rotational symmetries and the horizon has compact sections of non-toroidal topology.

Our results in 4d are unsurprising. We find that the only solution
is the near-horizon limit of the extremal Kerr black hole. In fact,
in the context of isolated horizons the same result has been
established~\cite{LP}. Observe that uniqueness of Kerr has only been
proved for non-extremal black holes; therefore our result can be
viewed as a first step towards proving uniqueness of extremal Kerr
among asymptotically flat black holes with degenerate horizons.
Pleasingly, our method in 4d worked just as easily with a negative
cosmological constant showing that the only regular solution is the
near-horizon geometry of extremal Kerr-AdS$_4$. It should be noted
that there are no known uniqueness theorems for asymptotically $AdS$
black holes even in 4d; perhaps our result will be useful in proving
uniqueness of extremal Kerr-AdS$_4$.

In five dimensions we were able to find all solutions in the pure
vacuum, i.e. zero cosmological constant. Naturally the results are
more complicated than in 4d. We found three families of near-horizon
geometries: two spherical topology horizons and one $S^1 \times S^2$
horizon. Further we identified how all the known vacuum extremal
black hole solutions fit into these families: i.e. extremal  boosted
Kerr string, extremal vacuum black ring, extremal Myers-Perry and
the extremal KK black holes (both slow and fast rotating). Our
results are summarised in detail in the Main results section. A
number of things may be deduced from our classification.

For example, one expects a vacuum doubly spinning black ring which is asymptotic to the KK monopole to exist (i.e. a ``Taub NUT'' black ring)\footnote{In fact a special case of this with one independent rotation parameter has been constructed~\cite{CEFGS}.}. Such a solution would have 4 parameters (roughly $J_i,M,P$). Presumably like other doubly spinning solutions in 5d it admits an extremal limit, which would be a three parameter family. One can then consider its near-horizon limit. From our Theorem 2, it follows that its near-horizon geometry is contained in our family of $S^1 \times S^2$ horizons. A reasonable guess is that it is simply given by the near-horizon limit of the extremal boosted Kerr string (which is a three parameter sub-family of our solution). The boost then would be related to the NUT parameter $P$ and as $P \to \infty$ (flat space limit) one must have $\sinh^2\beta \to 1$ in order to get the NH geometry of the asymptotically flat black ring, see~\cite{KLR}. In fact, for the asymptotically flat extremal black ring both the infinite radius limit and the near-horizon limit simplify to the tensionless (i.e. $\sinh^2\beta=1$) boosted Kerr string~\cite{FKLR}. In view of our near-horizon results it is thus natural to expect that the infinite radius limit of a KK black ring is the boosted Kerr string for arbitrary boost.

We also remark that a curious output of our analysis is that in some cases the near-horizon geometries we derived are isometric to the near-horizon limit of known black holes only in a subregion of parameter space. This occurs both for the $S^1\times S^2$ family and the second spherical topology case. It is possible these other regions of parameter space are occupied by unknown black hole solutions (e.g. KK black ring) but it seems more likely that such bounds on the parameters are invisible from the near-horizon geometry alone (e.g. as for the near-horizon of the asymptotically flat extremal ring which actually is only isometric to the tensionless boosted Kerr string in a subregion of its parameter space, see~\cite{FKLR}).

Other interesting consequences of our results regards uniqueness of
near-horizon geometries. Our analysis has revealed there are two
distinct classes of $S^3$ horizon geometries in 5d vacuum gravity.
Also the same near-horizon geometry can arise as the near-horizon
limit of different black holes although in all known examples the
black holes have different asymptotics (i.e KK or asymptotically
flat). Furthermore it seems clear that not all near-horizon
geometries arise as near-horizon limits of black holes with a given
asymptotics. For example, one can ask whether our second class of
$S^3$ topology horizon geometries can ever arise as the near-horizon
limit of an asymptotically flat extremal black hole. Due to its
$S^3$ topology one can identify the correct $U(1)$ generators which
must match onto those in the orthogonal 2-planes as asymptotic
infinity. One can therefore calculate the angular momenta via a
Komar integral over the horizon~\cite{FKLR} which gives \be
J_{\phi_1}= -\frac{4\pi \sqrt{2} \kappa^2}{G_5 C^8 \Gamma(\sigma_2)
\sqrt{C^2a_0-a_2c_2}}, \qquad J_{\phi_2}= -\frac{4\pi \sqrt{2}
\kappa^2}{G_5 C^8 \Gamma(\sigma_1) \sqrt{C^2a_0-a_2c_2}} \; . \ee It
is clear one can have $J_{\phi_1}=J_{\phi_2}$ (this occurs iff
$\Gamma(\sigma_2)=\Gamma(\sigma_1)$ which is equivalent to the
parameter $c_1=0$). Observe that the near-horizon geometry always
possesses exactly a $U(1)^2$ rotational symmetry group (i.e. it is
never enhanced even when $J_{\phi_1}=J_{\phi_2}$). However, from
group theoretic reasoning one might expect\footnote{In GR
kinematical arguments such as this are not sufficient to establish
symmetry enhancement; one usually uses dynamical input from the
Einstein's equation. In any case this symmetry enhancement occurs in
all known examples.} asymptotically flat black holes (with a single
horizon) with equal angular momenta to posses an enhanced rotational
symmetry group $SU(2)\times U(1)$ (recall the rotation group $SO(4)
\sim SU(2)\times SU(2)$). This leads us to conclude that this
near-horizon geometry does not correspond to that of an
asymptotically flat black hole. It should also be noted that in the
non-extremal case it has been shown~\cite{MI} that the Myers-Perry
black hole is the unique asymptotically flat black hole with two
rotational symmetries and $S^3$ topology horizon and one expects
this result to go over in the extremal case (and its near-horizon
geometry is in fact given by our other class of $S^3$ horizon
geometries).

Another useful aspect of this analysis is that the explicit metrics
for the various near-horizon geometries appear simple in the
coordinates we have derived. In contrast, the metrics one obtains by
taking the near-horizon limits of known solutions tend to be far
more complicated, as can be seen from the Appendix. This should make
the problem of generalizing our results to include gauge fields more
tractable. It would be interesting to classify the near-horizon
geometries of extremal, non-supersymmetric black holes in ungauged
supergravity theories. We intend to investigate this problem in the
near future.

One of the main motivations for this work was to investigate the existence of asymptotically AdS black rings. Unfortunately, we were not able to solve the vacuum near-horizon equations in the presence of a negative cosmological constant in general, even with the assumption of two rotational symmetries. This is in contrast to 4d where using the assumption of axisymmetry it was possible for us to do so. However, we did reduce the problem to solving a single 6th order ODE of one function. We found one set of solutions to this equation which correspond to the near-horizon geometry of the known, topologically $S^3$, extremal rotating AdS$_5$ black hole~\cite{HHT}. It would be interesting to find a solution which gives rise to the near-horizon geometry of an extremal AdS$_5$ black ring. By perturbing the near-horizon geometry of the asymptotically flat black ring we were able to construct an approximate near-horizon geometry corresponding to the near-horizon limit of a small (i.e. the size of the $S^1$ and $S^2$ are small compared to the AdS length scale) extremal black ring in AdS. The fact that the perturbation can always be made regular and preserves the $S^1 \times S^2$ topology appears to be non-trivial; perhaps this provides some evidence for the existence of, at least a small, extremal vacuum black ring in AdS$_5$.

\begin{center}{\bf{Acknowledgments}} \end{center}
HKK and JL are supported by STFC. We thank Pau Figueras, Mukund Rangamani and Harvey Reall for reading a draft of the manuscript and making useful comments. HKK would like to thank Eric Woolgar for helpful discussions concerning the black hole topology theorems.

\newpage
\appendix

\section{Global argument}
In this section we prove the following results quoted in the main text: $Q^2\mathcal{P}$ (needed in 4d) and $Q^3d^3\Gamma/d\sigma^3$ (needed in 5d) are globally defined functions on $\mathcal{H}$ which vanish where $Q$ vanishes.

This is not actually obvious as $\dot{\Gamma}$, $\ddot{\Gamma}$ and $d^3\Gamma/d\sigma^3$ need not be globally defined, although $\Gamma$ is.
To see this  note that the norm of $\partial / \partial \sigma$ is $\Gamma/Q$ which is regular everywhere except at the points where $Q$ vanishes. However, we know that $Q$ must vanish at two distinct points and thus we conclude that this vector field is not globally defined and thus $\frac{\partial}{\partial \sigma} \Gamma =\dot{\Gamma}$ and higher derivatives are not guaranteed to be globally defined. Note that this argument relies crucially on $Q$ vanishing somewhere. Recall this comes from the fact that $\sigma$ is a globally defined smooth non-constant\footnote{As dicusssed below~(\ref{Rickk}), $\sigma$ cannot be constant as otherwise $\det\gamma_{ij}=0$ everywhere} function on a compact space and thus $d\sigma$ vanishes at two distinct points (the max and min of $\sigma$). Then the invariant $(d\sigma)^2=Q/\Gamma$ tells us $Q\geq 0$ and vanishes at these two points.

To proceed we introduce the vector field $S= Q \frac{\partial}{\partial \sigma}$. Its norm squared is $\Gamma Q$ which is globally defined and vanishes at the zeroes of $Q$. $S$ is certainly regular everywhere except possibly at the zeros of $Q$. Let the zeroes of $Q$ be $\sigma_1<\sigma_2$. Then, assuming regularity, we have $Q=\dot{Q}_i(\sigma-\sigma_i) + \cdots $ near $\sigma=\sigma_i$ and since $Q \geq 0$ we learn that $\dot{Q}_1>0$ and $\dot{Q}_2<0$. This allows us to define $r_1^2=\sigma-\sigma_1$ and $r_2^2 =\sigma_2 -\sigma$. Then, near $\sigma_i$ we have $S \sim \frac{|\dot{Q_i}|}{2\Gamma_i} \; r_i \frac{\partial}{\partial r_i}$ which is regular at $r_i=0$ and vanishes there, as can be seen by using the Cartesian coordinates $x_i,y_i$ associated to $r_i$. We deduce that $S$ is a globally defined vector field on $\mathcal{H}$ which vanishes at the zeros of $Q$.

Thus we now employ the globally defined vector $S$ to construct invariants, e.g $S(\Gamma)=Q\dot{\Gamma}$ is globally defined (and vanishes where $Q$ does). Note the following identity:
\be
Q^2 \ddot{\Gamma} \equiv S(S(\Gamma)) -S(\Gamma)\dot{Q}
\ee
proves that $Q^2\ddot{\Gamma}$ is globally defined as $\dot{Q}$ must be (this is because $\dot{Q}$ is regular at the only potential problem points $\sigma=\sigma_i$ as $Q=\dot{Q}_i(\sigma-\sigma_i)+...$ ). Therefore $Q^2\ddot{\Gamma}$ is an invariant of the solution which vanishes at the zeros of $Q$ since $S$ vanishes at those points. Since $Q^2\mathcal{P}= 2Q^2\ddot{\Gamma}- S(\Gamma)^2/\Gamma- Q^2/\Gamma$ this proves that $Q^2\mathcal{P}$ is indeed globally defined and vanishes at the zeros of $Q$ . This establishes the result needed for the 4d analysis. The 5d case may be treated similarly using the identity
\be
Q^{3}\frac{d^3\Gamma}{d\sigma^3 }= S( Q^2 \ddot{\Gamma})- 2\dot{Q} Q^2\ddot{\Gamma}
\ee
which proves that $Q^3d^3\Gamma/d\sigma^3$ is globally defined (using the fact that $Q^2\ddot{\Gamma}$ is). Therefore $Q^3d^3\Gamma/d\sigma^3$ is an invariant which vanishes at the zeros of $Q$ as claimed.

\section{Near-horizon geometry of Kerr-AdS$_4$}
Use the form of the Kerr-AdS$_4$ metric as in~\cite{KP} which satisfies $R_{\mu\nu}=-3g^2 g_{\mu\nu}$ (our $g$ is the same as their $\alpha$). The angular velocity is given by $\Omega=a/(r_+^2+a^2)$ where $r_+$ is the largest zero of $\Delta_r = (r^2+a^2)(1+g^2r^2)-2mr$. Define \be
\rho^2= r^2+a^2 \cos^2\theta, \qquad \Delta_{\theta}=1-a^2g^2 \cos^2\theta, \qquad \Xi=1-g^2a^2 \; .
\ee
Using the algorithm  presented in~\cite{KLR} to determine the near-horizon data we find:
\bea
\label{kerradsdata}
k^{\Phi} &=& \frac{2ar_+}{(r_+^2+a^2)^2}, \qquad \Gamma=  \frac{\rho_+^2}{\Xi (r_+^2+a^2)}, \qquad A_0 = \frac{-\Delta_+''}{2\Xi (r_+^2+a^2)} \\
\gamma_{ab}dx^a dx^b &=& \frac{\rho_+^2}{\Delta_{\theta}} d\theta^2 + \frac{\sin^2\theta \Delta_{\theta} (r_+^2+a^2)^2}{\rho_+^2 \Xi^2} d\Phi^2 \; .
\eea
where $\Delta_+'' = (\Delta_r'')_{r=r_+}$ etc. Notice that in the flat space limit $g \to 0$ limit $r_+ \to a$ and thus the above near-horizon metric reduces correctly to that of Kerr as given in~\cite{KLR}.

\section{Near-horizon geometry of rotating AdS$_5$ black hole}
In this Appendix we present the near-horizon geometry of the known, topologically $S^3$, rotating AdS$_5$ black hole~\cite{HHT}.

\subsection{Self-dual case}
We first consider the self-dual case that occurs if the two independent angular momenta are set equal ($J_1 = J_2$). In this case the full solution exhibits symmetry-enhancement and it is convenient to treat it separately to the general case studied below. The self dual solution can be written in co-rotating
coordinates as: \be ds^2=- \frac{V(r)}{w(r)^2} dT^2+
\frac{dr^2}{V(r)} + \frac{r^2 w(r)^2}{4}[ d\psi+\cos\theta
d\phi-(\Omega(r)-\Omega_+)dT]^2 + \frac{r^2}{4}(
d\theta^2+\sin^2\theta d\phi^2) \ee where \be
V=1+g^2r^2-\frac{2M\Xi}{r^2}+ \frac{2Ma^2}{r^4}, \qquad
w(r)^2=1+\frac{2Ma^2}{r^4}, \qquad \Omega(r)= \frac{4Ma}{r^4w^2} \;
. \ee The horizon is located at the largest real root of $V(r)$,
$r=r_+$ so $V(r_+)=0$. Extremality implies $V'(r_+)=0$. The
near-horizon limit of this metric is given by the data \bea \Gamma
&=&
\frac{1}{w_+}, \qquad k^{\psi}=-\Omega_+' \qquad A_0=-\frac{V_+''}{2w_+}  \\
\gamma_{ab}dx^adx^b &=& \frac{r_+^2w_+^2}{4}(d\psi+\cos\theta
d\phi)^2+ \frac{r_+^2}{4}(d\theta^2+\sin^2\theta d\phi^2) \; .\eea

We will now show that the near-horizon metric
(\ref{NHdataGammaconst}) derived in the main text is identical to
the near-horizon of self-dual solution above. Consider
(\ref{NHdataGammaconst}) and define \be M \equiv \frac{4\Gamma
(C^2+2\Lambda \Gamma)^2}{(C^2+3\Lambda\Gamma)^3}, \qquad a^2 \equiv
\frac{2\Gamma(C^2+\Lambda\Gamma)}{(C^2+2\Lambda\Gamma)^2}, \qquad
r_+^2 \equiv \frac{4\Gamma}{C^2+3\Lambda\Gamma} \; \ee Observe that
these definitions imply $V(r_+)=0$, $V'(r_+)=0$,
$V''(r_+)=2C^2/\Gamma$ and that \be w_+^2 \equiv
w(r_+)^2=\frac{2(C^2+2\Lambda\Gamma)}{C^2+3\Lambda\Gamma} \ee where
$V$ and $w$ are defined as above. It now follows that the horizon
metric (\ref{NHdataGammaconst}) agrees exactly with that of self
dual solution. Now, using the definition of $\Omega$ above compute: \be
\Omega'(r_+)= -(C^2+3\Lambda\Gamma)\sqrt{
\frac{C^2+\Lambda\Gamma}{C^2+2\Lambda \Gamma}} \times \sqrt{
\frac{C^2+3\Lambda\Gamma}{2\Gamma^2(C^2+2\Lambda\Gamma)}} \; .\ee
Next, use the scaling freedom $\Gamma \to K \Gamma$ to set
$\Gamma=1/w_+$, which is equivalent to \be
\frac{C^2+3\Lambda\Gamma}{2\Gamma^2(C^2+2\Lambda\Gamma)}=1 \; .\ee
This then implies that in (\ref{NHdataGammaconst})
$k^{\psi}=-\Omega'(r_+)$ and $C^2=V''(r_+)/(2w_+)$ both of which
coincide with those for the self-dual solution given above. This completes the proof of
equivalence.

\subsection{General Angular Momenta}
 We now consider the general case for which the two independent angular momenta are not equal, i.e. $J_1 \neq J_2$. The solution satisfies $R_{\mu\nu} = -4g^2g_{\mu\nu}$ (we have set the parameter $l$ used in~\cite{HHT} to $g^{-1}$). The near horizon geometry is parameterized by three parameters $(r_+,a,b)$ subject to the extremality constraint
\begin{equation}
2g^2r_+^6 + r_+^4(1 + g^2b^2 + g^2a^2) - a^2b^2 = 0 \; .
\end{equation} Following the procedure given in~\cite{KLR}, it is straightforward to compute the near-horizon limit and we omit the details. The near-horizon metric can be written in the form~(\ref{canNH}) with horizon metric given by
\begin{equation}
\gamma_{ab}dx^a dx^b = \frac{\rho_+^2 d\theta^2}{\Delta_{\theta}} + \gamma_{ij}dx^{i}dx^{j}
\end{equation} with
\begin{eqnarray}
\gamma_{ij}dx^{i}dx^{j} &=& \frac{\Delta_{\theta}}{\rho_+^2}\left[\frac{(r_+^2+a^2)^2\sin^2\theta d\phi^2}{\Xi_a^2} + \frac{(r_+^2+b^2)^2\cos^2\theta d\psi^2}{\Xi_b^2}\right] \\ &+& \frac{1 + r_+^2g^2}{r_+^2\rho_+^2}\left[\frac{b(r_+^2+a^2)\sin^2\theta d\phi}{\Xi_a} + \frac{a(r_+^2+b^2)\cos^2\theta d\psi}{\Xi_b}\right]^2. \nonumber
\end{eqnarray} where
\begin{equation}
\Delta_{\theta} = 1 +g^2r_+^2 - g^2\rho_+^2 \qquad \rho_+^2 = r_+^2 + a^2\cos^2\theta + b^2\sin^2\theta \qquad \Xi_a = 1 - a^2g^2 \qquad \Xi_b = 1 - b^2g^2
\end{equation}
The remaining near-horizon data is
\begin{eqnarray}
\Gamma &=&  \frac{\rho_+^2 r_+^2}{(r_+^2+a^2)(r_+^2+b^2)}
 \qquad A_{0} = -\frac{4r_+^2(1+3g^2r_+^2 + g^2a^2+g^2b^2)}{(r_+^2+a^2)(r_+^2+b^2)} \nonumber \\
k^{\phi} &=& \frac{2a r_+ \Xi_a}{(r_+^2+a^2)^2} \qquad k^{\psi} =
\frac{2b r_+ \Xi_b}{(r_+^2+b^2)^2} \; .
\end{eqnarray}
Note that the above formulas simplify in the zero cosmological
constant case $g=0$ -- in particular $r_+^2=|ab|$. We should also
note that there is no loss of generality in assuming $a>b>0$.

\section{Near-horizon geometry of KK black hole}
In this section we give the near-horizon geometries of the extremal KK black holes found in~\cite{Rasheed} (see also~\cite{Larsen}). We will use the form of the solution as given in~\cite{EM}. The non-extremal solution carries the 4d conserved charges $(M,Q,P,J)$ (i.e. it a rotating dyonic black hole) and we will choose an orientation for rotation such that $J\geq 0$. In 5d when $P \neq 0$ it has horizon topology $S^3$ and is asymptotic to the KK monopole. When $P=0$ it is merely the boosted Kerr-string and thus we only consider the $P \neq 0$ case in this section. As is well known there are two different extremal limits of this black hole called slowly rotating (since $G_4J<PQ$) and fast rotating (since $G_4J>PQ$).

\subsection{Slowly rotating solution}
This extremal limit of the KK black hole, is given by $a,m \to 0$ with $\eta =a /m<1$ fixed. This extremal solution can be parameterized by three positive constants $(p,q,\eta)$. In this case the angular velocities are:
\be
\Omega_y = \sqrt{ \frac{p+q}{q}}, \qquad \Omega_{\phi}=0 \; .
\ee
After some calculation one can show that the near-horizon is of the form (\ref{canNH}) with the metric on $H$ given by
\be
\gamma_{ab}dx^adx^b = H_p d\theta^2 +\frac{H_q}{H_p}(dy+A_{\phi}d\phi)^2 + \frac{(pq)^3(1-\eta^2) \sin^2\theta d\phi^2}{4(p+q)^2 H_q}
\ee
where
\be
H_p= \frac{p^2q}{2(p+q)} (1+\eta \cos\theta), \qquad H_q= \frac{pq^2}{2(p+q)} (1-\eta \cos\theta), \qquad A_{\phi}= \frac{q^2 p^{5/2}}{2(p+q)^{3/2} H_q}(\eta-\cos\theta)
\ee
and regularity of the horizon demands $y\sim y+ 8\pi P$ (or quotients) and $\phi \sim \phi+2\pi$ where $P= \sqrt{\frac{p^3}{4(p+q)}}$ and $0 \leq \theta \leq \pi$. Coordinates which are adapted to the $U(1)^2$ rotational symmetry can be defined by $\phi=\phi_1+\phi_2$ and $y=2P(\phi_2-\phi_1)$; absence of conical singularities then implies $\phi_1,\phi_2$ are $2\pi$ periodic with $\partial/ \partial \phi_1$ vanishing at $\theta=\pi$ and $\partial /\partial \phi_2$ vanishing at $\theta=0$ -- i.e. one must have $S^3$ topology. The other near-horizon data is
\be
A_0=- \frac{2(p+q)}{(pq)^{3/2}(1-\eta^2)^{1/2}}, \qquad \Gamma=\frac{2(p+q)}{(pq)^{3/2}(1-\eta^2)^{1/2}} H_p
\ee
and
\be
k^{\phi}= -\frac{2(p+q)\eta}{(pq)^{3/2}(1-\eta^2)}, \qquad k^y= \frac{2}{1-\eta^2} \sqrt{ \frac{p+q}{q^3}} \;.
\ee

There is a special case which simplifies considerably, $\eta=0$ (note this gives $J=0$). Defining $y= p\sqrt{p/(p+q)} \psi$ one gets:
\be
\gamma_{ab}dx^a dx^b= \frac{p^2q}{2(p+q)} [ d\theta^2+\sin^2\theta d\phi^2 + 2(d\psi-\cos\theta d\phi)^2 ]
\ee
and
\be
\Gamma= \sqrt{ \frac{p}{q}}, \qquad C^2 = \frac{2(p+q)}{(pq)^{3/2}}
\ee
and
\be
k= \frac{2(p+q)}{(pq)^{3/2}} \frac{\partial}{\partial \psi}= C^2\frac{\partial}{\partial \psi} \; .
\ee
Noting that
\be
\Gamma C^{-2}= \frac{p^2q}{2(p+q)}
\ee
it is easy to see this is of the form of the $\Gamma=a_0$ case we derived in the main text. To prove complete equivalence one needs to invert the parameter change which is easily done:
\be
p=C^{-1}\sqrt{2\Gamma(1+\Gamma^2)}, \qquad q= C^{-1} \sqrt{ \frac{2(1+\Gamma^2)}{\Gamma^3}} \; .
\ee

\subsection{Fast rotating solution}
This extremal limit of the KK black hole, is given by $m=a>0$. This extremal solution can be parameterized by three positive constants $(p,q,a)$ which satisfy $p,q \geq 2a$. In this case the angular velocities are:
\be
\Omega_y = \sqrt{\frac{{(q^2-4a^2)}}{q(p+q)}}, \qquad \Omega_{\phi}= \frac{1}{\sqrt{pq}}
\ee
After some calculation one can show that the near-horizon is of the form (\ref{canNH}) with the metric on $H$ given by
\be
\gamma_{ab}dx^adx^b = H_p d\theta^2 +\frac{H_q}{H_p}(dy+A_{\phi}d\phi)^2 +\frac{pqa^2 \sin^2\theta}{H_q} d\Phi^2
\ee
where
\be
\label{Hfast}
H_p = -a^2 \sin^2\theta +\frac{p(pq+4a^2)}{2(p+q)}+ \frac{2pQP}{\sqrt{pq}} \cos\theta , \quad H_q=-a^2 \sin^2\theta +\frac{q(pq+4a^2)}{2(p+q)}- \frac{2qQP}{\sqrt{pq}} \cos\theta
\ee
and
\be
A_{\phi}= -\frac{2P}{H_q} (H_q+a^2 \sin^2\theta)\cos\theta+ \sqrt{\frac{p}{q}} \, \frac{Q(2a^2(p+q)+q(p^2-4a^2)) \sin^2\theta}{(p+q)H_q}
\ee
and
\be
P= \sqrt{\frac{p(p^2-4a^2)}{4(p+q)}}, \qquad Q= \sqrt{\frac{q(q^2-4a^2)}{4(p+q)}} \; .\ee
Regularity of the horizon demands $y\sim y+ 8\pi P$ (or quotients) and $\phi \sim \phi+2\pi$. Coordinates which are adapted to the $U(1)^2$ rotational symmetry can be defined by $\phi=\phi_1+\phi_2$ and $y=2P(\phi_2-\phi_1)$; absence of conical singularities then implies $\phi_1,\phi_2$ are $2\pi$ periodic with $\partial/ \partial \phi_1$ vanishing at $\theta=\pi$ and $\partial /\partial \phi_2$ vanishing at $\theta=0$ -- i.e. one must have $S^3$ topology. The other near-horizon data is
\be
A_0=-\frac{1}{a \sqrt{pq} }, \qquad \Gamma= \frac{H_p}{a \sqrt{pq}}
\ee
and
\be
k^{\phi}= \frac{pq+4a^2}{2a^2 \sqrt{pq} (p+q)}, \qquad k^y= -\frac{(p^2-4a^2)Q}{qa^2(p+q)} \; .
\ee

\section{$D=5, \Lambda =0$ special cases}
In this appendix we provide details concerning special cases arising in the $\Lambda =0$ and $\Gamma = a_0 + a_2\sigma^2$ case analysed in the main text.
\subsection{Exclusion of a special case}
In this subsection we show that the case
\be
c_1^2a_0a_2+(C^2a_0+a_2c_2)^2 = 0
\ee
is not compatible with having a compact horizon. Observe that this case implies
that the polynomial $P(\sigma)= \alpha \sigma^2 + \beta \sigma + \gamma$ has vanishing discriminant.  From~(\ref{pos}), $\dot\omega =0$ and we may shift $x^1$ to set $\omega =0$. Since $(C^2a_0 + a_2c_2) = \pm c_1\sqrt{-a_0a_2}$, it follows
\begin{equation}
\gamma_{11} = \frac{2\alpha(\sigma-\sigma_0)^2}{\Gamma},
\end{equation} $\alpha = \mp a_2c_1\sqrt{-a_0a_2}$ and
\begin{equation}
\sigma_0 = \pm \left(-\frac{a_0}{a_2}\right)^{1/2} \; .
\end{equation} Further, $\Gamma = a_2(\sigma-\sigma_0)(\sigma+\sigma_0)$ and hence the horizon metric is
\begin{equation}\label{D=0hor}
\gamma_{ab}dx^{a}dx^{b} = \frac{\Gamma d\sigma^2}{Q} + \frac{2\alpha(\sigma-\sigma_0)(dx^1)^2}{a_2(\sigma+\sigma_0)} + \frac{Q(dx^2)^2}{2\alpha(\sigma-\sigma_0)^2}.
\end{equation} Having obtained the local form of the horizon metric, we turn to its regularity. The roots of $Q$ in this case are easily seen to be
\begin{equation}
\sigma_{\pm} = \frac{c_1}{2C^2} \pm \frac{C^2a_0 - a_2c_2}{2C^2\sqrt{-a_0a_2}}.
\end{equation} Now suppose $\sigma_0 >0$; then it is easy to show $\sigma_+ =\sigma_0$. Similarly $\sigma_0 < 0$ implies $\sigma_- = \sigma_0$. Therefore in either case, $(d\sigma)^2 = Q/\Gamma$ vanishes only at one point. This implies that~(\ref{D=0hor}) cannot describe a compact manifold and hence we exclude this case.

\subsection{$\alpha = 0$}
Consider now the special case $\alpha=0$. Note that since $\alpha=0$, $a_0=-a_2c_2C^{-2}$, which implies $\gamma=0$ and therefore $\beta \neq 0$. Observe that another way of
writing the solution to (\ref{pos}), valid when $\beta \neq 0$ (and any $\alpha$) is
\be \omega= \pm \left[ \frac{ \kappa(a_2\sigma^2 -a_0) }{\beta
P(\sigma)} +c_3' \right] \; . \ee The advantage of this expression
is that it is valid when $\alpha=0$ and it is related to
(\ref{omegaflat}) by $c_3=c_3'+(\kappa a_2)/(\beta \alpha)$. Thus,
setting $\alpha=0$ gives \be \omega= \pm \left[
\frac{C}{4\sqrt{c_1^2c_2^2a_2}} \frac{C^2\sigma^2+ c_2}{a_2
\sigma}+c_3' \right] \ee and $a_2>0$. Also note that
$\Gamma=a_2(\sigma^2-c_2C^{-2})$ and \be
\gamma_{11}=\frac{4a_0a_2c_1 \sigma}{\Gamma}=4a_0a_2 \left(
\frac{Q(\sigma)}{\Gamma}+C^2 \right) \ee so $a_0>0$ and thus $c_2<0$
. We must have $c_1 \sigma>0$ and without loss of
generality we choose $\sigma>0$ so $c_1>0$. Therefore
$\gamma_{11}>0$ for $\sigma_1\leq \sigma \leq \sigma_2$ and hence
the horizon metric is non-degenerate everywhere except at the points
$\sigma_i$ where there are conical singularities. The rest of the
analysis is identical to the $\alpha \neq 0$ case and one obtains
the same values for $d_i$ noting that $P(\sigma_i)=2a_0a_2c_1
\sigma_i$.

\end{document}